\PassOptionsToPackage{unicode}{hyperref}
\PassOptionsToPackage{hyphens}{url}
\PassOptionsToPackage{dvipsnames,svgnames,x11names}{xcolor}
\documentclass[
  12pt]{article}

\usepackage{amsmath,amssymb}
\usepackage{iftex}
\ifPDFTeX
  \usepackage[T1]{fontenc}
  \usepackage[utf8]{inputenc}
  \usepackage{textcomp} 
\else 
  \usepackage{unicode-math}
  \defaultfontfeatures{Scale=MatchLowercase}
  \defaultfontfeatures[\rmfamily]{Ligatures=TeX,Scale=1}
\fi
\usepackage{lmodern}
\ifPDFTeX\else  
\fi
\IfFileExists{upquote.sty}{\usepackage{upquote}}{}
\IfFileExists{microtype.sty}{
  \usepackage[]{microtype}
  \UseMicrotypeSet[protrusion]{basicmath} 
}{}
\makeatletter
\@ifundefined{KOMAClassName}{
  \IfFileExists{parskip.sty}{%
    \usepackage{parskip}
  }{
    \setlength{\parindent}{0pt}
    \setlength{\parskip}{6pt plus 2pt minus 1pt}}
}{
  \KOMAoptions{parskip=half}}
\makeatother
\usepackage{xcolor}
\makeatletter
\ifx\paragraph\undefined\else
  \let\oldparagraph\paragraph
  \renewcommand{\paragraph}{
    \@ifstar
      \xxxParagraphStar
      \xxxParagraphNoStar
  }
  \newcommand{\xxxParagraphStar}[1]{\oldparagraph*{#1}\mbox{}}
  \newcommand{\xxxParagraphNoStar}[1]{\oldparagraph{#1}\mbox{}}
\fi
\ifx\subparagraph\undefined\else
  \let\oldsubparagraph\subparagraph
  \renewcommand{\subparagraph}{
    \@ifstar
      \xxxSubParagraphStar
      \xxxSubParagraphNoStar
  }
  \newcommand{\xxxSubParagraphStar}[1]{\oldsubparagraph*{#1}\mbox{}}
  \newcommand{\xxxSubParagraphNoStar}[1]{\oldsubparagraph{#1}\mbox{}}
\fi
\makeatother

\usepackage{longtable,booktabs,array}
\usepackage{calc} 
\usepackage{etoolbox}
\makeatletter
\patchcmd\longtable{\par}{\if@noskipsec\mbox{}\fi\par}{}{}
\makeatother
\IfFileExists{footnotehyper.sty}{\usepackage{footnotehyper}}{\usepackage{footnote}}
\makesavenoteenv{longtable}
\usepackage{graphicx}
\makeatletter
\def\maxwidth{\ifdim\Gin@nat@width>\linewidth\linewidth\else\Gin@nat@width\fi}
\def\maxheight{\ifdim\Gin@nat@height>\textheight\textheight\else\Gin@nat@height\fi}
\makeatother
\setkeys{Gin}{width=\maxwidth,height=\maxheight,keepaspectratio}
\makeatletter
\def\fps@figure{htbp}
\makeatother

\makeatletter
\@ifpackageloaded{caption}{}{\usepackage{caption}}
\AtBeginDocument{%
\ifdefined\contentsname
  \renewcommand*\contentsname{Table of contents}
\else
  \newcommand\contentsname{Table of contents}
\fi
\ifdefined\listfigurename
  \renewcommand*\listfigurename{List of Figures}
\else
  \newcommand\listfigurename{List of Figures}
\fi
\ifdefined\listtablename
  \renewcommand*\listtablename{List of Tables}
\else
  \newcommand\listtablename{List of Tables}
\fi
\ifdefined\figurename
  \renewcommand*\figurename{Figure}
\else
  \newcommand\figurename{Figure}
\fi
\ifdefined\tablename
  \renewcommand*\tablename{Table}
\else
  \newcommand\tablename{Table}
\fi
}
\@ifpackageloaded{float}{}{\usepackage{float}}
\floatstyle{ruled}
\@ifundefined{c@chapter}{\newfloat{codelisting}{h}{lop}}{\newfloat{codelisting}{h}{lop}[chapter]}
\floatname{codelisting}{Listing}

\makeatother
\makeatletter
\@ifpackageloaded{caption}{}{\usepackage{caption}}
\@ifpackageloaded{subcaption}{}{\usepackage{subcaption}}
\makeatother

\ifLuaTeX
  \usepackage{selnolig}  
\fi
\usepackage[]{natbib}
\usepackage{bookmark}

\IfFileExists{xurl.sty}{\usepackage{xurl}}{} 
\hypersetup{
  pdftitle={Simultaneous Inference for Watermark Proportions in LLM-Generated Text},
  pdfkeywords={Closed testing; post-selection inference; simultaneous confidence bounds; structure-adaptive inference; text watermarking},
  colorlinks=true,
  linkcolor={blue},
  filecolor={Maroon},
  citecolor={Blue},
  urlcolor={Blue},
  pdfcreator={LaTeX}}

\newcommand{\anon}{1}

\if1\anon
  \hypersetup{pdfauthor={Jingcheng He; Zinan Zhao; Wenguang Sun; Xintao Xia}}
\fi

\usepackage{amsthm,bm,enumerate}
\usepackage{algorithm,algpseudocode}
\usepackage{flafter,placeins}
\algrenewcommand\algorithmicrequire{\textbf{Input:}}
\algrenewcommand\algorithmicensure{\textbf{Output:}}

\newtheorem{theorem}{Theorem}
\newtheorem{lemma}{Lemma}

\newtheorem{proposition}{Proposition}

\theoremstyle{definition}

\newtheorem{assumption}{Assumption}

\newtheorem{property}[theorem]{Property}

\newcommand{\E}{\operatorname{\mathbb{E}}} 

\providecommand{\ind}{\mathbf{1}}

\providecommand{\FDP}{\operatorname{FDP}}

\newcommand{\Nset}{\mathcal{N}}
\newcommand{\TDP}{\operatorname{TDP}}
\newcommand{\positivepart}[1]{[#1]_+}

\begin{document}

\def\spacingset#1{\renewcommand{\baselinestretch}%
{#1}\small\normalsize} \spacingset{1}


\newcommand{\Title}{Simultaneous Inference for Watermark Proportions in LLM-Generated Text}

\if1\anon
{
  \title{\bf \Title}
  \author{
Jingcheng He\thanks{School of Mathematical Sciences and Center for Data Science, Zhejiang University. First author.}
\and
Zinan Zhao\thanks{Department of Statistics, University of Michigan}
\and
Wenguang Sun\thanks{Center for Data Science, Zhejiang University}
\and
Xintao Xia\thanks{Center for Data Science, Zhejiang University; Email: \texttt{xintaox@zju.edu.cn}. Corresponding author.}
}

  \date{\today}
  \maketitle
}
\fi

\if0\anon
{
  \bigskip
  \bigskip
  \bigskip
  \begin{center}
    {\LARGE\bf \Title}
  \end{center}
  \medskip
} \fi

\bigskip
\begin{abstract}
Watermarking enables detection of text generated by large language models, but practical review also requires quantifying retained signals within content of interest. Readers may choose passages based on their content and refine their selections after examining the watermark evidence. We develop simultaneous lower confidence bounds on the fraction of selected tokens that retain a watermark signal. The bounds hold jointly for the full document and any passages selected during review. We develop structure-adaptive closed testing (SACT), which uses a working model of sequential signal clustering to construct scores for calibrated sign-based tests. Under a conditional null model, the bounds have finite-sample simultaneous validity even when the working model is misspecified. An exact reference-family representation recovers the full closed-testing bounds while enabling efficient repeated queries. In a constructed database-recovery review, SACT detects document-wide watermark signals and provides lower confidence bounds on watermark proportions in selected passages. Assessing related passages together gives a sharper lower bound on their combined signal count. In simulations spanning two watermarking mechanisms and a range of signal strengths and densities, SACT generally has higher power than competing methods.

\end{abstract}

\noindent%
{\it Keywords:} Closed testing; post-selection inference; simultaneous confidence bounds; structure-adaptive inference; text watermarking.
\vfill

\newpage
\spacingset{1.8} 

\section{Introduction}\label{sec:introduction}

Documents drafted with the assistance of large language models (LLMs) can interleave human-written and generated passages \citep{lee2022coauthor}. In scientific and professional writing \citep{noy2023productivity,liang2025quantifying}, the distinction can matter when a reader evaluates the reasoning behind a claim or recommendation. For example, a database-recovery review may combine excerpts from a published outage report with LLM-generated analyses of maintenance decisions, backup failures, and recovery procedures. Evidence of LLM generation somewhere in the review does not establish whether the model contributed to the wording of a particular explanation or recommendation. A reader may therefore ask how much of the document was generated by the model and whether generated wording appears in the passages supporting a decision.

Text watermarking provides evidence for these assessments by embedding a statistical signal during generation that a verifier can detect using a secret key \citep{kirchenbauer2023watermark,kuditipudi2024robust,dathathri2024scalable,li2026textgrain}. A positive global test provides evidence of a watermark but does not quantify how much of the document contains the signal or identify the passages in which it occurs \citep{li2025framework,li2026robust}. Editing may disrupt the signal in generated text \citep{li2026robust}. The fraction of tokens retaining watermark evidence therefore provides a conservative measure of the watermarked model's contribution. Recent work estimates watermark proportions in mixed-source documents \citep{li2025proportion}, identifies watermarked segments \citep{li2024segmenting,zhao2025segments,bonnerjee2026segmentation}, and selects individual token positions with evidence of watermarking \citep{blanchet2026localization}. Two features of practical review motivate our work.

First, the passages of interest may change during review. A reader examining a recommendation for a recovery exercise may return to an earlier explanation of a backup failure and decide to assess the two excerpts together. When watermark evidence guides these choices, a \textit{post-selection inference} problem arises: a bound calibrated for a prespecified passage may overstate the watermark proportion if it is applied to passages chosen for their strong evidence. Recent watermarking methods account for selection through data fission \citep{wei2026fissionguard} or simultaneous bounds on historical watermark matches \citep{chen2026countscan}. We aim to assess selected passages jointly and provide a lower confidence bound on the fraction of tokens that retain the original watermark signal, even as readers refine their selections of passages.

Second, using token order can improve detection power and sharpen lower confidence bounds on watermark proportions. Watermark signals tend to cluster within generated passages, so evidence from neighboring tokens can help distinguish a watermarked passage from isolated large scores. Giving greater weight to positions within signal clusters can also help preserve power when editing replaces some watermarked tokens or adds unwatermarked text. The challenge is to exploit this sequential structure while preserving simultaneous validity for any set of tokens a reader selects.

We develop a framework for simultaneous inference on watermark proportions. It provides lower confidence bounds that hold jointly over all nonempty sets of token positions. To construct these bounds, we develop \emph{structure-adaptive closed testing} (SACT), which uses a working model to learn sequential structure from the magnitudes of transformed token-level statistics and reserves their signs for testing. Under the conditional null model, the bounds remain valid even if the working model is misspecified. We also derive an exact reference-family representation of the closed-testing bounds. After fitting the working model and calibrating the local tests, we compute the reference family once per document. We then reuse it to evaluate the bounds efficiently as readers select different passages.

In a constructed database-recovery review, SACT detects document-level watermark signals and quantifies their extent in passages of interest. The bounds hold simultaneously for all selections, so readers can explore the report interactively, choosing passages as questions arise.

\subsection{Contributions}

Our contributions are as follows.

\begin{enumerate}

    \item \emph{A framework for interactive watermark inference.} We provide simultaneous lower confidence bounds on watermark proportions over all sets of token positions. These bounds remain valid as readers refine their selections in light of the text and watermark evidence. The framework encompasses global detection, document- and passage-level proportion inference, and token-level discovery.

    \item \emph{Structure-adaptive closed testing with an exact computational shortcut.} We develop SACT to exploit sequential signal clustering through structure-adaptive token-level scores. We derive an exact reference-family representation that separates document-level preprocessing from computing bounds for selected token sets. After model fitting and calibration, this representation allows the full closed-testing bounds for $M$ queries on a document of $n$ tokens to be computed in $O(n^2+Mn)$ time.

    \item \emph{Theoretical guarantees for structure-adaptive inference.} Under Assumption~\ref{ass:null-pivots}, we establish finite-sample simultaneous coverage even when the working model is misspecified, allowing SACT to exploit sequential structure while preserving inferential validity. 
\end{enumerate}

\subsection{Related work}

Watermarking embeds detectable signals in LLM-generated text. Green-list methods modify token probabilities to favor a keyed subset of the vocabulary \citep{kirchenbauer2023watermark}, potentially affecting text quality. Unbiased watermarking preserves the model's next-token distribution when averaged over fresh watermark randomness \citep{hu2024unbiased,kuditipudi2024robust}. Our formulation focuses on unbiased schemes, including Gumbel--max sampling and the tournament sampling used by SynthID-Text \citep{dathathri2024scalable}.

\paragraph*{Watermark detection and proportion inference.}
Recent work develops document-level watermark tests from token-level pivotal statistics, providing a framework for statistically efficient detection \citep{li2025framework}. Related methods account for dependence induced by pseudorandom collisions \citep{cai2025collision} and use truncated goodness-of-fit tests to improve robustness to human edits \citep{li2026robust}. Beyond detecting watermark presence, mixture models based on pivotal statistics enable estimation of watermark proportions in hybrid texts \citep{li2025proportion}. Applied to the full document, our framework yields a lower confidence bound on the document-level retained-signal proportion. A positive bound corresponds to rejecting the global null hypothesis of no watermark signal.

\paragraph*{Localization in mixed-source text.}
Identifying which parts of a document retain watermark signals is important for practical use and has motivated a range of localization methods. Window and multiscale scans identify regions with concentrated watermark evidence \citep{kirchenbauer2024reliability,pan2025waterseeker,zhao2025segments,sander2026textseal}, while change-point methods estimate segment boundaries \citep{li2024segmenting,li2025adaptive_segmentation,bonnerjee2026segmentation}. Token-level methods provide asymptotic marginal FDR control for selected positions \citep{blanchet2026localization}. FissionGuard separates passage selection from validation and controls the false discovery rate among reported passages \citep{wei2026fissionguard}. SACT provides a different guarantee: a simultaneous lower bound on the retained-signal proportion that remains valid even when readers use the same watermark scores to select passages and refine their choices.

\paragraph*{Simultaneous inference and structure adaptation.}
Our approach draws on exploratory multiple-testing methods to construct simultaneous confidence bounds that remain valid under data-dependent selection \citep{goeman2011exploratory,goeman2019simultaneous}. CountScan \citep{chen2026countscan} bounds historical context--token matches simultaneously over intervals and extends these bounds to arbitrary sets by aggregating bounds for disjoint intervals. SACT targets tokens that retain their original watermark signal and uses structure-adaptive closed testing to bound their proportion directly over the selected set. Closed testing \citep{marcus1976closed,goeman2011exploratory} converts intersection tests into simultaneous true-discovery bounds. Within the framework of \citet{goeman2021admissible}, admissible coherent simultaneous procedures admit a closed-testing representation. Related methods use reference families \citep{blanchard2020reference} or FDP envelopes along rejection paths \citep{hemerik2019permutation,katsevich2020simultaneous}; jointly valid reference-family constraints can be interpolated to arbitrary sets. Direct computation of the full closure requires exponentially many intersection tests. For Simes tests, shortcuts give the bound efficiently for any selected set \citep{goeman2019simultaneous}. FACT uses symmetry and monotonicity of the local tests to reduce the computation to polynomial time \citep{dobriban2020fact}. Our work is also related to methods that use structural information in multiple testing \citep{lei2018adapt,lei2021star} and conformal inference \citep{zhao2025plis}. In closed testing, the power of the local tests determines how informative the resulting bounds are. Related constructions include KJI \citep{li2024knockofftdp}, which uses supplied coin-flip scores, and ACODE \citep{magnani2026collective}, which learns conformity scores calibrated using reference observations. Motivated by these developments, we propose SACT to incorporate sequential signal structure into simultaneous inference over arbitrary token sets while preserving validity under working-model misspecification. We further derive an exact reference-family representation that makes the full closed-testing bounds efficiently computable.

The remainder of the paper is organized as follows. Section~\ref{sec:problem} introduces the watermark evidence and inferential target. Section~\ref{sec:method} develops SACT through a database-review application and summarizes supporting numerical studies. Section~\ref{sec:theory} gives the theoretical guarantees, reference-family representation, and exact shortcut. Section~\ref{sec:con} summarizes the findings and outlines directions for future work.

\section{Watermark Analysis: Questions and Framework}\label{sec:problem}

Section~\ref{subsec:review-questions} introduces the questions that arise in document-wide assessment and interactive content review. Section~\ref{subsec:watermark-evidence} defines the retained-signal proportion and states the simultaneous coverage guarantee. Section~\ref{subsec:framework-overview} outlines how SACT constructs the bounds.

\subsection{Application context and scientific questions}\label{subsec:review-questions}

Documents drafted jointly by people and LLMs can contain both human-written and model-generated passages \citep{lee2022coauthor}. Knowing where a model contributed matters because its suggestions can shape the views expressed in the final text \citep{jakesch2023cowriting}. A reader may assess the document as a whole or select passages for closer inspection as the review proceeds. These two forms of review motivate the following questions:
\begin{enumerate}
    \item \emph{Document-wide assessment.} Does the document contain LLM-generated text, and if so, to what extent?
    \item \emph{Interactive content review.} How much of the text in the passages a reader selects during review was generated by an LLM?
\end{enumerate}
The first question requires both detecting LLM involvement and quantifying the proportion of generated text. For the second question, conclusions must remain valid as reviewers choose passages for their content and refine their selections in light of the evidence. A further challenge is that individual tokens may provide little evidence of LLM involvement. When generated text occurs in contiguous passages, combining evidence from neighboring tokens can make the signal easier to detect.

\subsection{Simultaneous inference for watermark proportions}
\label{subsec:watermark-evidence}

Watermarking provides evidence for the assessments in Section~\ref{subsec:review-questions} by embedding a statistical signal in generated text during token sampling \citep{kirchenbauer2023watermark,kuditipudi2024robust,dathathri2024scalable}. Given the watermark protocol and secret key, a verifier computes a statistic at each token position to quantify evidence for this signal \citep{li2025framework,li2026robust}.

Let $W_{1:n}=(W_1,\ldots,W_n)$ denote the observed, possibly edited document of $n$ tokens, which may represent words or word fragments. For $j\in[n]=\{1,\ldots,n\}$, let $Y_j$ denote the verification statistic at position $j$, with larger values indicating stronger watermark evidence. Following the edit model of \citet{li2026robust}, let $\xi_j\in\{0,1\}$ indicate whether position $j$ contains a watermark signal. We set $\xi_j=1$ if $W_j$ is unchanged from a watermarked generation step and its observed seeding context matches the one used at that step; otherwise, $\xi_j=0$. An edit to the preceding context can therefore disrupt the signal even when $W_j$ is unchanged. The null hypotheses and the set of true null positions are
\begin{equation}\label{eq:elementary-null}
    H_j:\xi_j=0,\quad j\in[n],
    \qquad
    \Nset=\{j\in[n]:\xi_j=0\}.
\end{equation}
Under Assumption~\ref{ass:null-pivots}, $Y_j$ follows a known distribution $F_0$ whenever $\xi_j=0$.

Let $R\subseteq[n]$ be a nonempty set of token positions selected for review. The true discovery count $d(R)$ and true discovery proportion (TDP) $\TDP(R)$ are, respectively, the number and fraction of tokens in $R$ that retain a watermark signal:
\begin{equation}\label{eq:tdp-target}
    d(R)=\sum_{j\in R}\xi_j=|R\setminus\Nset|,
    \qquad
    \TDP(R)=\frac{d(R)}{|R|}.
\end{equation}
By definition, $\TDP(R)$ is also the watermark proportion in $R$. 
A token that retains the signal must come from the watermarked model, but a token from that model can lose its signal after editing. A lower confidence bound on $\TDP(R)$ therefore also bounds from below the fraction of tokens in $R$ that came from the model.
However, a confidence bound calibrated for a prespecified set need not retain its nominal coverage if $R$ is chosen after examining the text and its watermark evidence. To address both review questions, we construct simultaneous lower confidence bounds on $\TDP(R)$, drawing on confidence envelopes for the false discovery proportion (FDP) \citep{genovese2006exceedance} and closed testing for exploratory inference \citep{goeman2011exploratory,goeman2021admissible}. For a prescribed $\alpha\in(0,1)$, we seek lower bounds $\underline{\TDP}_\alpha(R)$ satisfying
\begin{equation}\label{eq:simultaneous-goal}
    \Pr\big\{\underline{\TDP}_\alpha(R)\leq\TDP(R)
    \text{ for every nonempty }R\subseteq[n]\big\}\geq 1-\alpha.
\end{equation}
The corresponding lower confidence bound on $d(R)$ is $d_\alpha(R)=|R|\underline{\TDP}_\alpha(R)$, which we call the \emph{certified count}. Different choices of $R$ address the two review questions.

\begin{enumerate}
    \item \emph{Document-wide assessment.} For $R=[n]$, the bound gives a minimum fraction of tokens in the document that retain a watermark signal. A positive bound also provides evidence of watermark presence, linking global detection \citep{li2025framework,li2026robust} to document-level proportion inference \citep{li2025proportion}.

    \item \emph{Interactive content review.} Readers can define $R$ using token-level watermark evidence, the content of the passages under review, or both.

    At the token level, readers can vary a threshold $t$ and examine where the selected positions $R_t=\{j:Y_j\geq t\}$ cluster in the text \citep{blanchet2026localization}. For each nonempty $R_t$, $1-\underline{\TDP}_\alpha(R_t)$ gives an upper confidence bound on the realized FDP, $\FDP(R_t)=1-\TDP(R_t)$. For a desired FDP level $\gamma$, readers may choose any $t$ for which $R_t$ is nonempty and the upper bound is at most $\gamma$.

    At the content level, $R$ may correspond to a single passage, as in passage-level localization \citep{li2024segmenting,zhao2025segments,bonnerjee2026segmentation}, or include sentences from several passages that support the same claim or explanation. The bound applies to the selected text as a whole and remains valid as readers refine their selections after examining the evidence.
\end{enumerate}
Furthermore, $R$ can take more flexible forms tailored to task-specific goals and constraints, including sets obtained through algorithmic search, as illustrated in Section~\ref{subsec:database-exploratory}. For any nonempty set $\widehat R$ selected or revised after examining the text and watermark evidence, Equation~\eqref{eq:simultaneous-goal} gives $\Pr\{\underline{\TDP}_\alpha(\widehat R)\leq\TDP(\widehat R)\}\geq 1-\alpha$ without further adjustment for selection.

\subsection{A framework for interactive watermark analysis}\label{subsec:framework-overview}

\begin{figure}[!htbp]
\centering
\includegraphics[width=\linewidth]{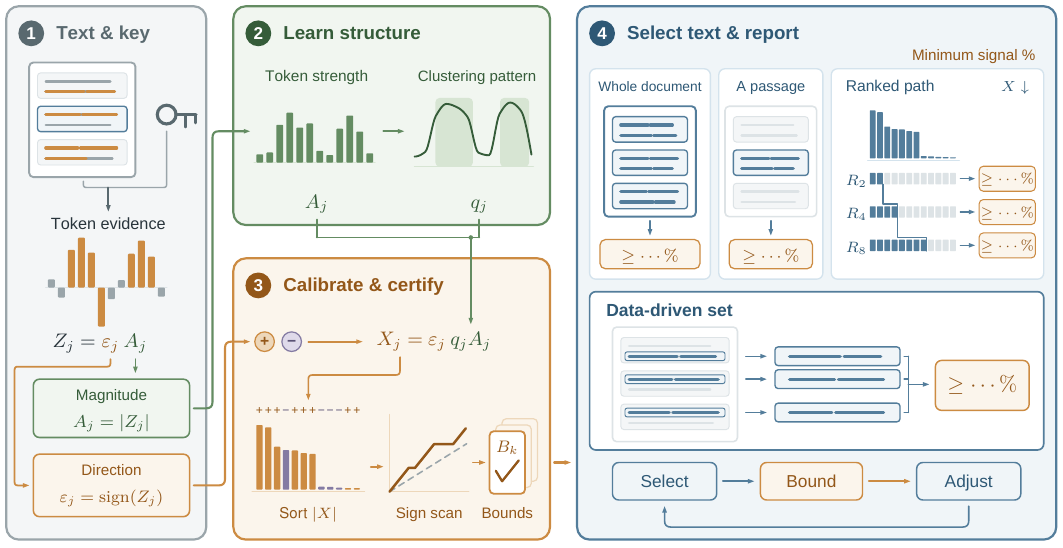}
\caption{Overview of structure-adaptive closed testing (SACT). Score magnitudes guide model fitting, and the signs provide evidence for local tests. Closed testing yields simultaneous lower confidence bounds on the signal proportion in any selected set of tokens.}
\label{fig:wm-procedure}
\end{figure}

Closed testing provides a classical construction of the bounds in Equation~\eqref{eq:simultaneous-goal}. SACT combines this construction with structure-adaptive scoring and an exact computational shortcut. Figure~\ref{fig:wm-procedure} illustrates the overall workflow, and we outline the steps below.

\begin{enumerate}
    \item \emph{Structure-adaptive scoring.} Watermark signals can cluster within generated passages \citep{li2024segmenting,zhao2025segments,bonnerjee2026segmentation}. We fit a working model to the magnitudes of transformed token-level statistics, reserving their signs for testing. The fitted weight for each token takes neighboring observations into account. We multiply the transformed statistic by this weight to obtain the structure-adaptive score used for testing.

    \item \emph{Sign calibration and closed testing.} Under the conditional null model, the fitted scores satisfy sign-flip symmetry (Property~\ref{prop:sign-flip}). This property permits calibration of sign-based local tests even when the working model is misspecified \citep{barber2015knockoffs,lei2018adapt,zhao2025plis}. Closed testing combines these tests into simultaneous lower confidence bounds on $d(R)$ and $\TDP(R)$ for every nonempty $R\subseteq[n]$ \citep{marcus1976closed,goeman2011exploratory}.

    \item \emph{Efficient evaluation of selected sets.} Building on reference-family ideas \citep{blanchard2020reference,katsevich2020simultaneous}, we derive an exact representation of the SACT bounds that avoids enumerating all intersection tests. The reference family is computed once for the document and reused to recover the full closed-testing bound for each new selected set.
\end{enumerate}

\section{Application to a Database-Recovery Review}\label{sec:method}

We apply the framework in Section~\ref{subsec:framework-overview} to a constructed database-recovery review, using watermark evidence to assess the extent of LLM involvement in the document and in content selected during interactive review.

\subsection{Document construction and review questions}\label{subsec:application-data}

We constructed a database-recovery review for a hypothetical service team using attributed excerpts from the GitLab postmortem on the January 2017 outage.\footnote{Source: \href{https://about.gitlab.com/blog/postmortem-of-database-outage-of-january-31/}{GitLab, \emph{Postmortem of database outage of January 31}}, February 10, 2017.} We organized the review around three topics: safeguards for destructive maintenance operations, backup restoration checks, and recovery drills.

For each topic, we used Qwen3.5-9B \citep{qwen2026qwen35} to draft an analysis and placed it after the corresponding source excerpts. The target lengths were 512, 768, and 1,024 tokens for the maintenance, backup, and recovery analyses, respectively. We used Gumbel--max watermarking \citep{kuditipudi2024robust} with full-vocabulary sampling at temperature 1 and a watermark seed based on the four preceding tokens. To mimic the effects of human editing, we then used the same Qwen3.5-9B model with watermarking disabled to revise the wording subject to a $3\%$ limit on token edits. The final document contains $5,454$ tokens, of which $2,095$ lie in the generated and subsequently edited sections.

\subsection{Watermark generation and verification}\label{subsec:application-verification}

During watermarked generation, the secret key and the four tokens preceding the current position determine the watermark seed. From this seed, the decoder generates one pseudorandom value for each candidate token. The Gumbel--max rule combines these values with the next-token probabilities from the LLM to select the next token.

Let $W_{1:n}$ denote the final report, with $n=5,454$ tokens. Using the same tokenizer, key, and seeding rule as in generation, the verifier reconstructs the watermark values from the observed context. Let $U_{j,w}$ denote the reconstructed value for vocabulary token $w$ at the position $j$. For the observed token $W_j$, the verification statistic and its $p$-value are
\begin{equation*}
    Y_j=U_{j,W_j},\qquad p_j=1-Y_j.
\end{equation*}
Larger $Y_j$, or equivalently smaller $p_j$, indicates stronger watermark evidence. Verification does not require access to the next-token probabilities used during generation.

Repeated four-token contexts reuse watermark values, so we filter out all but the first token position for each context. This gives a set $\mathcal E$ of eligible positions, with $n_e=|\mathcal E|=5{,}223$. During fitting, we treat the statistics at the remaining $231$ positions as missing but preserve their locations in the sequence. For a selected set $R$, only eligible positions can contribute to the certified count $d_\alpha(R)$, whereas the denominator $|R|$ includes all selected tokens. Among the eligible positions, $2{,}008$ retain watermark signals.

\subsection{Constructing structure-adaptive scores}
\label{subsec:watermark-strength}

To prioritize tokens for review, we weight watermark scores using evidence from neighboring positions. We learn the weights from score magnitudes and reserve the signs for testing.

\paragraph*{Masking and working emissions.}
Let $F_0$ denote the null distribution function of the verification statistic $Y_j$. Under Assumption~\ref{ass:null-pivots}, the eligible null statistics are independent draws from $F_0$ conditional on the eligible positions, the signal indicators, and the nonnull statistics. When $F_0$ is continuous, we can transform $Y_j$ at each eligible position $j\in\mathcal E$ to a normal score and separate its sign from its magnitude:
\begin{equation*}
    Z_j=\Phi^{-1}\{F_0(Y_j)\}=\varepsilon_jA_j,
    \qquad
    \varepsilon_j=\operatorname{sign}(Z_j),
    \qquad
    A_j=|Z_j|,
\end{equation*}
where $\Phi$ is the standard normal distribution function. Under the same conditioning, the null scores are independent standard normal variables. After further conditioning on their magnitudes, their signs remain independent and equally likely to be positive or negative.

The database review uses Gumbel--max watermarking. In the background state ($G_j=0$), the pivot $Y_j$ is uniform, so $F_0(y)=y$ on $[0,1]$. Equation~\eqref{eq:wm-gumbel-alternative} gives the pivot distribution for a token generated with this watermark. It is a mixture of beta distributions in which a token with next-token probability $p>0$ contributes a $\mathrm{Beta}(1/p,1)$ component with weight $p$. On the $-\log Y_j$ scale, each component is exponential with mean $p$. The verifier generally does not know these probabilities. Appendix~\ref{app:gamma-mixture} shows how a continuous approximation to the mixture leads to the following one-parameter working model for the active state ($G_j=1$),
\begin{equation}\label{eq:gamma-pivot-emission}
    -\log Y_j\mid G_j=1\sim\mathrm{Gamma}(\nu,1),
    \qquad 0<\nu\leq1.
\end{equation}
At $\nu=1$, $-\log Y_j$ is exponential with mean one, so $Y_j$ is uniform as in the background state. Smaller values of $\nu$ place more probability on pivots near one.

The corresponding density of $A_j$ in the background state, $g_0$, is standard half-normal. To obtain the active-state density $g_1$ at $a>0$, we add the densities of $Z_j$ at $a$ and $-a$. The resulting densities are given in Equation~\eqref{eq:hmm-working-emissions}. For SynthID-Text, we use the centered binomial score defined in Appendix~\ref{app:synthid}. Under the null, the score is equally likely to be positive or negative given its magnitude.

\paragraph*{Modeling signal clustering.}
Our main working model for signal clustering is a two-state hidden Markov model (HMM), which we use in the database review. In the simulations, we also consider a change-point model. Both models use the same densities $g_0$ and $g_1$ for the magnitudes at eligible positions and treat observations at the remaining positions as missing.

\begin{itemize}
    \item \emph{Hidden Markov model.}
The HMM for $G_{1:n}$ borrows information across neighboring positions without requiring prespecified passage boundaries. The active state $G_j=1$ can persist across a passage even when individual observations provide weak evidence, as may occur after editing or a change in seeding context. We place a continuous prior on the transition parameters and average the forward--backward state probabilities over their posterior distribution to obtain the weight $q_j=\widehat{\Pr}(G_j=1\mid A_{\mathcal E})$. Appendix~\ref{app:hmm-fitting} describes the prior and model fitting.

    \item \emph{Change-point model.}
Motivated by change-point \citep{li2024segmenting} and epidemic-segment \citep{bonnerjee2026segmentation} approaches to watermark localization, our change-point (CP) working model represents the active region $\mathcal I$ as a union of $K$ disjoint intervals, with $G_j=\mathbf{1}\{j\in\mathcal I\}$. We assign a truncated Poisson prior to $K$ and, conditional on $K$, give equal prior probability to all admissible interval configurations. We use dynamic programming to average over the number and locations of the intervals, yielding the weight $q_j=\widehat{\Pr}(j\in\mathcal I\mid A_{\mathcal E})$. Appendix~\ref{app:cp-fitting} gives the prior specification and recursions.
\end{itemize}

Under either working model, $q_j$ is the fitted probability that position $j$ belongs to an active region. This use of a posterior probability is related to the local index of significance \citep{sun2009dependent}. We use $q_j$ to weight the score $Z_j$, giving
\begin{equation}\label{eq:sign-flip-score}
    X_j=q_jZ_j=\varepsilon_jq_jA_j,
    \qquad j\in\mathcal E.
\end{equation}
The magnitude $|X_j|=q_jA_j$ gives priority to positions supported by both the individual observation and evidence from neighboring positions. In the simulations in Section~\ref{sec:numerical}, SACT is typically more powerful with the combined score $q_jZ_j$ than with either $Z_j$ or $\varepsilon_jq_j$ (Figure~\ref{fig:sact-app-paths-2048}). Because the weights are positive and depend only on magnitudes, changing a null sign leaves the fitted weights and score magnitudes unchanged. Under the conditional null model, the weighted scores therefore retain sign-flip symmetry even if the working model for clustering is misspecified. Section~\ref{subsec:score-symmetry} establishes this symmetry \citep{barber2015knockoffs}, which we use to calibrate the local tests in Section~\ref{subsec:sign-scan-test}.

\begin{figure}[!htb]
\centering
\includegraphics[width=\linewidth]{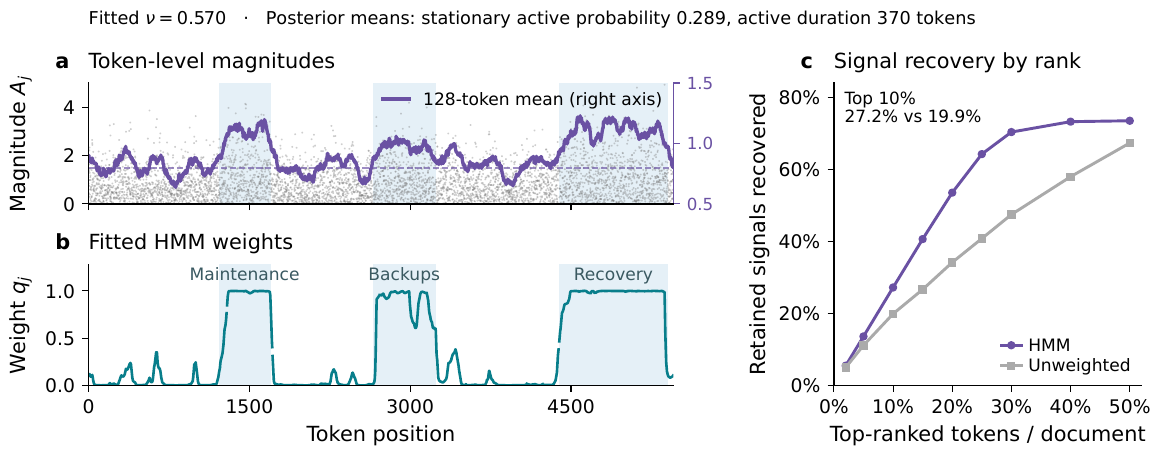}
\caption{Structure-adaptive scoring in the database review. \textbf{a}, Eligible token magnitudes (left axis) and their moving mean over centered 128-token windows (purple, right axis), excluding ineligible positions. The dashed line marks the half-normal null mean on the right axis. \textbf{b}, HMM weights fitted to the magnitudes. \textbf{c}, Fraction of recorded retained signals captured by the HMM-weighted (decreasing $X_j=q_jZ_j$) and unweighted (decreasing $Z_j$) rankings. Shading marks the three passages on maintenance, backup, and recovery that the LLM drafted and later revised.}
\label{fig:database-evidence}
\end{figure}

\paragraph*{Model fitting and diagnostics.}
For the database report, the HMM fit gives $\widehat\nu=0.570$. The posterior means of the stationary active-state probability and active-run duration are 0.289 and 370 tokens, respectively. Figure~\ref{fig:database-evidence} shows how the fit changes the ranking. Among eligible positions, the median weight is 0.995 within the generated-and-edited analyses and 0.006 in the surrounding material. At a selection size of 10\% of document tokens, ranking by decreasing $X_j=q_jZ_j$ captures 27.2\% of the recorded retained signals, compared with 19.9\% when ranking directly by decreasing $Z_j$.

To assess whether the HMM exploits sequential clustering, we randomly permute the observed magnitudes across eligible positions and refit the model. The gain in masked log-likelihood over the all-null model was 185.0 in the original order, compared with 114.1--121.5 across 64 shuffled orders (median 115.9). Thus, the HMM captured structure in the ordering of the magnitudes beyond their marginal distribution. Assumption~\ref{ass:null-pivots} implies that the null pivots are uniform and that the normal scores at adjacent null positions are uncorrelated. In this report, the Kolmogorov--Smirnov distance between the empirical distribution of the known-null pivots and $\mathrm{Uniform}(0,1)$ was 0.012. The sample correlation between the unweighted normal scores $Z_j$ at adjacent null positions was 0.0045. Both findings are consistent with the assumption.

\subsection{Calibrating tests and computing bounds}
\label{subsec:sign-scan-test}

We construct simultaneous lower confidence bounds through closed testing, which requires a local test of signal absence for every subset of eligible positions $\mathcal E$. Drawing on the scanning idea of higher criticism \citep{donoho2004higher}, we use a sign scan to accommodate strong watermark signals at a few token positions as well as weaker signals spread across more positions.

For a nonempty set $I\subseteq\mathcal E$ of size $m$, let $H_I$ denote the hypothesis that no position in $I$ retains a watermark signal. Let $i_1,\ldots,i_m$ be the positions in $I$ ordered by decreasing score magnitude, so that $|X_{i_1}|\geq\cdots\geq|X_{i_m}|$, with ties resolved by a fixed rule independent of the signs. This magnitude ordering reserves the signs for testing, whereas the signed-score ranking in Figure~\ref{fig:database-evidence} prioritizes tokens for review. Specifically, we test $H_I$ using
\begin{equation}\label{eq:sign-scan-statistic}
    T_I=\max_{1\leq k\leq m}
         \frac{\sum_{\ell=1}^k\varepsilon_{i_\ell}}{\sqrt{k}},
    \qquad T_\varnothing=-\infty.
\end{equation}
For each $k$, the numerator is the excess of positive over negative signs among the $k$ highest-ranked positions. Under $H_I$, Property~\ref{prop:sign-flip} implies that, conditional on the score magnitudes, the ordered signs are independent and equally likely to be positive or negative. We use this joint null distribution to calibrate the maximum over all $k$. The local test $\phi_I^{\mathrm{scan}}=\ind\{T_I\geq c_{\alpha,m}\}$ uses a cutoff that depends only on $m$ and $\alpha$. In this review, we use $\alpha=0.05$ and exact cutoffs with null rejection probability at most $\alpha$. Appendix~\ref{app:sign-tests} describes their computation.

The sign-scan test above corresponds to the $s=2$ member of the power-divergence family in Equation~\eqref{eq:power-divergence-sign-scan}, hence the name SACT-$2$. We use it as our default because its statistic has a simple form based on cumulative sums of the ordered signs. Appendices~\ref{app:performance-comparisons} and~\ref{app:sact-additional-results} compare it with other choices of $s$ and alternative local tests.

Applying the closed-testing procedure \citep{marcus1976closed,goeman2011exploratory} to these local tests gives the following lower confidence bound on the signal count in any selected set $R$:
\begin{equation}\label{eq:database-count-bound}
    d_\alpha(R)=\min_{I\subseteq\mathcal E:\,\phi_I^{\mathrm{scan}}=0} |(R\cap\mathcal E)\setminus I|.
\end{equation}
To see why the minimum gives a lower bound, consider the set of all eligible null positions. With probability at least $1-\alpha$, its local test does not reject, so this set is among the candidates $I$ in Equation~\eqref{eq:database-count-bound}. For this particular $I$, every position in $(R\cap\mathcal E)\setminus I$ retains a signal. The minimum over all candidates is therefore at most the true signal count in $R$. Because the same event gives the bound for every $R$, the bounds hold simultaneously, as stated in Theorem~\ref{thm:sact-validity}. For nonempty $R$, the proportion lower bound is $\underline{\TDP}_\alpha(R)=d_\alpha(R)/|R|$.

Direct evaluation of Equation~\eqref{eq:database-count-bound} would require searching over all subsets of $\mathcal E$. To avoid this search, we rank the scores by magnitude and use the resulting nested sets to compute the bound. Section~\ref{subsec:reference-family} gives the exact reference-family representation.

Specifically, sort the positions once by decreasing $|X_j|$, and let $S_k$ contain the positive-score positions among the first $k$ ranks, with $S_0=\varnothing$. We construct lower bounds $\ell_k$ on the retained-signal counts in $S_k$ that hold jointly with probability at least $1-\alpha$.
To apply these bounds to a passage $R$, consider where the signals counted by $\ell_k$ can lie. At most $|S_k\setminus R|$ of them can lie outside the passage. The rest must lie in $R$, and we can take the largest such lower bound over all $k$. The same argument applies to any selected set, including a union of passages, giving
\begin{equation}\label{eq:database-interpolation}
    d_\alpha(R)=\max_{0\leq k\leq n_e}\bigl(\ell_k-|S_k\setminus R|\bigr)_+,
    \qquad (x)_+=\max\{x,0\}.
\end{equation}
For nonzero scores with distinct magnitudes, this calculation recovers the full closed-testing bound. For each $R$, the proof of Theorem~\ref{thm:computational-bound} constructs a set $I$ whose local test does not reject and for which $|(R\cap\mathcal E)\setminus I|$ equals the count in Equation~\eqref{eq:database-interpolation}. After fitting and calibration, constructing the reference sets $S_k$ and their lower bounds $\ell_k$ for a document of $n$ tokens takes $O(n^2)$ time, and evaluating the bound for each selected set takes $O(n)$ time. The total cost for $M$ sets is therefore $O(n^2+Mn)$. Algorithm~\ref{alg:sact-bound} in Appendix~\ref{app:proofs-computation} gives the preprocessing and query steps.

\subsection{Findings across review targets}\label{subsec:application-review}

We first assess watermark presence and extent in the full report ($R=[n]$; Question~1), then examine where the evidence concentrates using three types of review targets: sets of highly ranked tokens, sliding windows, and sentence-aligned passages (Question~2). To assess the contribution of structural weighting, we compare SACT-$2$ with its unweighted counterpart ($X_j=Z_j$). The two procedures use the same local test and significance level and are evaluated on the same windows and selected passages, with token-level selections based on their respective score rankings.

For token-level selection, let $R_k$ contain the $k$ eligible positions with the largest signed scores $X_j$. For each FDP target $\gamma\in\{0.1,0.2,0.5\}$, we select the largest $R_k$ satisfying $\underline{\TDP}_\alpha(R_k)\geq 1-\gamma$, or the empty set if none qualifies. For passage-level assessment, we evaluate sliding windows of 128, 256, and 512 tokens, advancing each window by 16 tokens. To obtain readable excerpts, we also maximize the TDP lower bound over passages of consecutive complete sentences containing at least 64 tokens, subject to the review budget. For each procedure, the bounds hold simultaneously across all these selections.

\begin{figure}[!htb]
\centering
\includegraphics[width=\linewidth]{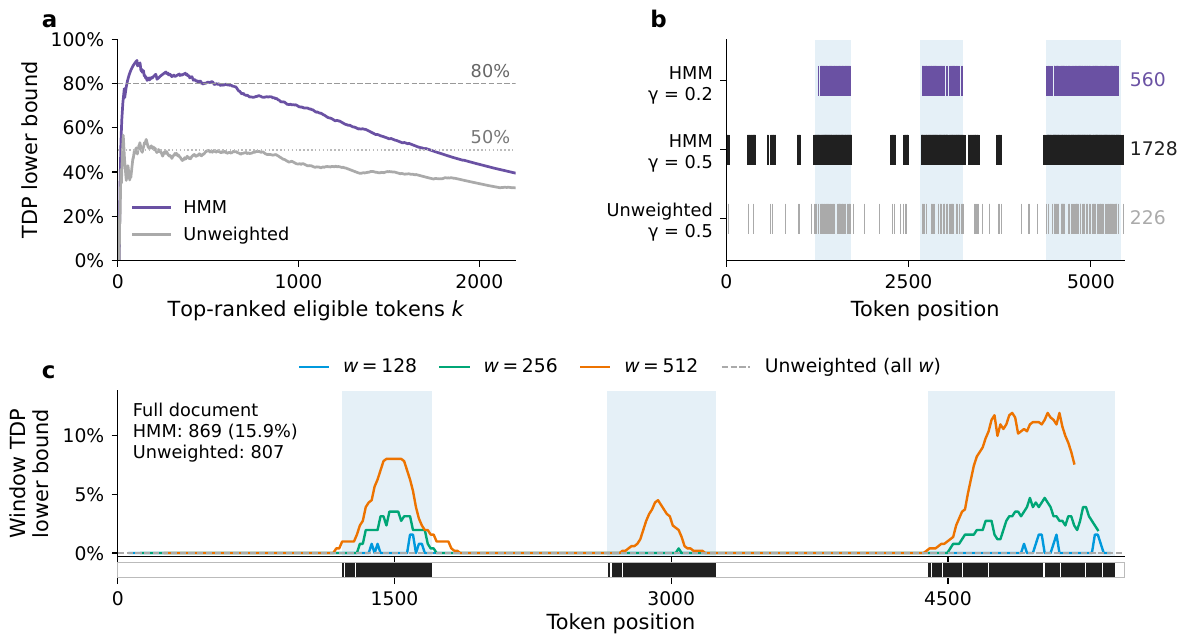}
\caption{Token-level discovery and window localization. \textbf{a}, Lower confidence bounds on the signal proportion among the top $k$ ranked tokens. \textbf{b}, Selected token positions at the indicated FDP targets, with counts shown on the right. \textbf{c}, Proportion lower bounds for sliding windows. The strip below marks recorded retained signals.}
\label{fig:database-localization}
\end{figure}

\paragraph*{Global detection and proportion.}
For the full report ($R=[n]$), SACT-$2$ certifies that at least 869 of the 5,454 tokens retain the watermark signal, giving a lower confidence bound of 15.9\% (Figure~\ref{fig:database-localization}(c)). The unweighted procedure certifies at least 807 signals, or 14.8\%. Both procedures detect watermark signals, with weighting giving a slightly higher global lower bound. For comparison, the 2,008 eligible tokens known to retain the signal account for 36.8\% of the full report.

\paragraph*{Token-level discovery.}
Token-level selection complements the document-wide assessment by highlighting where watermark evidence concentrates. At the FDP target $\gamma=0.2$, the HMM-weighted procedure selects 560 positions and certifies that at least 448 of them retain the watermark signal (Figure~\ref{fig:database-localization}(a) and (b)). The unweighted procedure yields an empty selection at this target. The generation record shows that all 560 selected positions retain the signal. Relaxing the target to $\gamma=0.5$ expands the HMM-weighted selection to 1,728 positions, compared with 226 under unweighted scoring. Structure adaptation thus yields a much larger gain in token-level discovery than in the document-wide lower bound. The selected positions cluster near the maintenance, backup, and recovery analyses, directing closer review toward these parts of the report.

\paragraph*{Passage-level localization.}
We next localize watermark evidence directly using sliding windows. The SACT-$2$ bounds peak near the maintenance, backup, and recovery analyses (Figure~\ref{fig:database-localization}(c)). The largest lower confidence bound on the signal proportion increases from 1.6\% for 128-token windows to 4.7\% for 256-token windows and 11.9\% for 512-token windows. Larger windows can accumulate evidence that is difficult to certify at finer resolutions. The unweighted procedure gives zero lower bounds for every window examined.

To obtain readable excerpts, we also consider passages of consecutive complete sentences, with lengths constrained by a reading budget. For each budget $B$, we select the passage of at most $B$ tokens that maximizes the lower confidence bound on the signal proportion. At budgets of 256, 512, and 1,024 tokens, SACT-$2$ certifies at least 12 retained signals among 251 tokens (4.8\%), 54 among 489 (11.0\%), and 220 among 994 (22.1\%), respectively (Figure~\ref{fig:database-findings}(a) and (b)). All three excerpts lie in or around the recovery analysis and cover restore exercises, dependencies, and assumptions about recovery time. On these same three excerpts, unweighted scoring certifies only 0, 0, and 2 retained signals. Although the full-report bounds were similar (15.9\% with SACT-$2$ and 14.8\% with unweighted scoring), SACT-$2$ certified substantially more retained signals in each of the three recovery passages.

\subsection{Joint assessment of passages under a reading budget}\label{subsec:database-exploratory}

The preceding analysis selected a single passage for each reading budget. A reader assessing a proposed restore exercise may also want to review earlier passages on maintenance and backup procedures. Under a fixed reading budget, the reader can select relevant passages from all three analyses. To illustrate this form of review, we use an algorithm to select passages and assess watermark evidence jointly across the selected text.

Specifically, let $\mathcal C_{B,m}$ be the collection of nonempty sets formed by taking the union of at most $m$ disjoint passages. Each passage consists of complete sentences and contains at least 64 tokens; the combined length is at most $B$ tokens. We seek to maximize $\underline{\TDP}_\alpha(R)$ over $R\in\mathcal C_{B,m}$, with $B\in\{256,512,1024,1536,2048\}$ and $m\in\{1,2,3\}$. For $m=1$, we enumerate all feasible passages. For $m>1$, we use the heuristic in Appendix~\ref{app:passage-search} to combine candidate passages and refine their sentence boundaries to improve the joint lower bound.

\begin{figure}[!h]
\centering
\includegraphics[width=\linewidth]{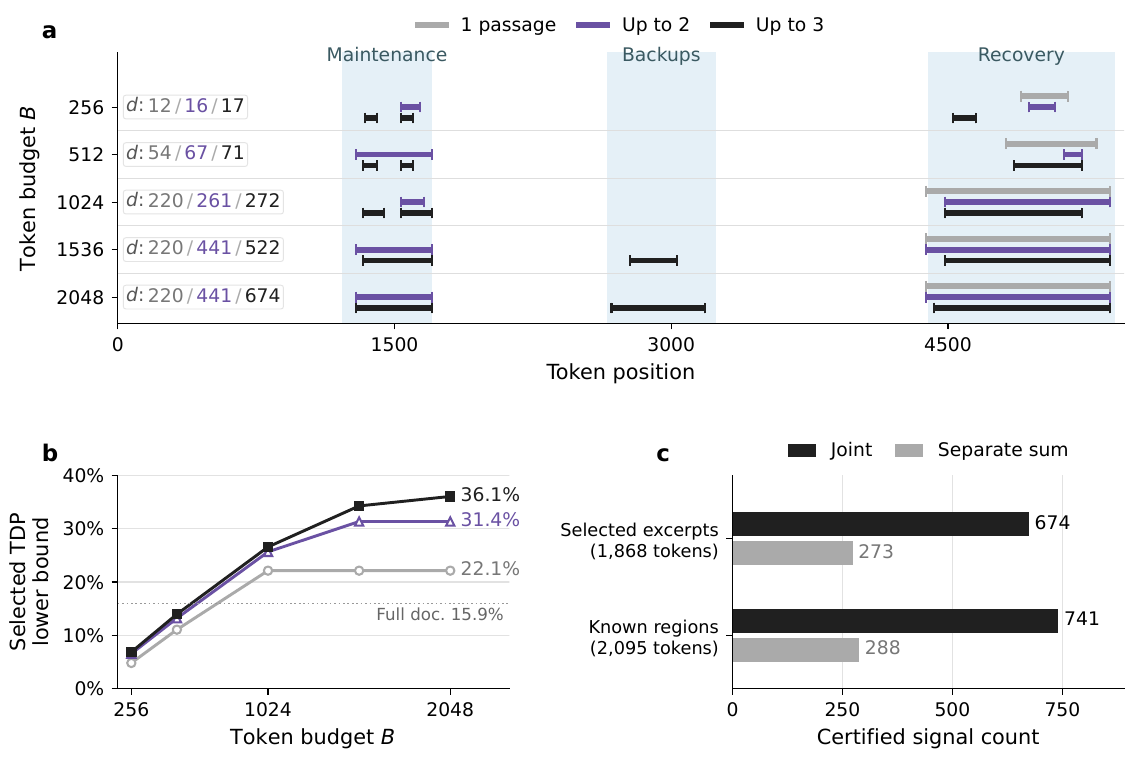}
\caption{Joint assessment of sentence-aligned passages under reading budgets $B$ and passage limits $m$. \textbf{a}, Selected passage locations and joint certified counts, with counts listed in legend order. \textbf{b}, Lower confidence bounds on the retained-signal proportion in the selected text. The dotted line marks the full-document bound. \textbf{c}, Joint certified counts versus sums of counts certified separately for the same passages, shown for the selected passages at $B=2048$, $m=3$ and the three complete analyses with known boundaries. The three shaded passages are those the LLM drafted and later revised. Labels indicate their topics.}
\label{fig:database-findings}
\end{figure}

The selected passages vary with the reading budget and passage limit (Figure~\ref{fig:database-findings}(a) and (b)). At $B=1536$ and $m=3$, the selected excerpts cover all three topics and jointly certify at least 522 retained signals among 1,522 tokens (34.3\%). At $B=2048$ and $m=3$, selecting three passages reduces the text to review from 5,454 to 1,868 tokens. SACT certifies at least 674 retained signals in these passages (36.1\%), compared with 869 in the full report. According to the generation record, 97.5\% of the tokens in the selected passages retain the watermark signal. Focusing on these passages substantially reduces the amount of text to review while capturing most of the retained watermark signals in the report.

To see how much joint assessment contributes, we evaluate the same three passages separately (Figure~\ref{fig:database-findings}(c)). With the same fitted scores and calibrated tests, the lower bounds for the individual passages are 41, 23, and 209. Their sum certifies 273 retained signals, or 14.6\% of the combined text, compared with 674 signals (36.1\%) when assessing the union directly. Pooling evidence across the three passages gives a sharper lower bound on their overall watermark proportion, and reading them together shows how the LLM contributed to the report's explanations of the failures and its recommendations.

Within the budget and passage limit, the search seeks a high joint bound. Reviewers can also select passages by content and revise their choices after seeing the evidence. Simultaneous coverage applies to either selection rule \citep{goeman2011exploratory}.

\subsection{Supporting numerical studies}\label{sec:numerical}

To further assess the performance of SACT, we compare it with alternative simultaneous inference procedures under varying signal proportions and strengths. We draw background documents from the Colossal Clean Crawled Corpus (C4), a collection of cleaned English web text from Common Crawl \citep{raffel2020t5}. In each document, we replace two randomly selected, separated intervals of $L$ tokens with equal-length passages generated by Qwen3.5-9B \citep{qwen2026qwen35} under Gumbel--max watermarking with four-token local-context seeding.

We vary document length $n$, passage length $L$, and sampling temperature $t$ (100 replicates per setting). Methods share verification statistics, eligible positions, windows, and ranking paths. See Table~\ref{tab:sact-central} for a representative setting and Appendices~\ref{app:technical-details} and~\ref{app:sact-experiments} for details.

\begin{table}[!htbp]\centering
\renewcommand{\baselinestretch}{1}\fontsize{8}{10}\selectfont
\setlength{\tabcolsep}{4.5pt}
\begin{tabular}{@{}lcccccccc@{}}\toprule
Method & \shortstack{\strut HMM/CP\\[-3pt]\strut weighting} & \shortstack{\strut Global\\[-3pt]\strut power} & \shortstack{\strut Certified\\[-3pt]\strut count} & \shortstack{\strut Window\\[-3pt]\strut count} & $|R_Y|$ & \shortstack{\strut $|R_X|$\\[-3pt]\strut HMM} & \shortstack{\strut $|R_X|$\\[-3pt]\strut CP} & CPU (s)\\\midrule
SACT-1 & \textcolor{ForestGreen}{$\checkmark$} & 0.950 [0.887, 0.984] & 66.6 (3.4) & 26.1 (1.5) & 25.2 (3.8) & 112.0 (7.2) & 104.7 (7.3) & 0.033\\
SACT-1.5 & \textcolor{ForestGreen}{$\checkmark$} & \textbf{0.960} [0.901, 0.989] & 67.8 (3.4) & 26.4 (1.5) & 22.0 (3.8) & 114.2 (7.3) & 105.6 (7.5) & 0.034\\
SACT-2 & \textcolor{ForestGreen}{$\checkmark$} & \textbf{0.960} [0.901, 0.989] & 68.4 (3.4) & 26.7 (1.5) & 18.8 (3.7) & \textbf{114.7} (7.4) & 106.5 (7.5) & 0.033\\
SACT-2-CP & \textcolor{ForestGreen}{$\checkmark$} & \textbf{0.960} [0.901, 0.989] & \textbf{69.2} (3.4) & 27.2 (1.5) & 21.1 (3.9) & 109.5 (7.8) & \textbf{119.2} (7.5) & 0.033\\
\midrule
SACT-BC & \textcolor{ForestGreen}{$\checkmark$} & 0.950 [0.887, 0.984] & 33.4 (2.0) & 17.7 (1.2) & 27.2 (3.6) & 62.8 (4.1) & 59.9 (4.3) & 0.032\\
SACT-KJI-B & \textcolor{ForestGreen}{$\checkmark$} & 0.950 [0.887, 0.984] & 48.7 (2.7) & 22.9 (1.3) & \textbf{29.5} (3.8) & 87.2 (5.7) & 82.6 (5.8) & 0.018\\
SACT-KJI-D & \textcolor{ForestGreen}{$\checkmark$} & 0.950 [0.887, 0.984] & 46.9 (2.7) & 21.4 (1.3) & \textbf{29.5} (3.8) & 82.2 (5.6) & 77.6 (5.6) & 0.014\\
\midrule
SACT-$q$ & \textcolor{ForestGreen}{$\checkmark$} & \textbf{0.960} [0.901, 0.989] & 68.2 (3.4) & 25.7 (1.6) & 0.0 (0.0) & 25.8 (6.5) & 23.7 (6.5) & 0.032\\
SACT-$Z$ & \textcolor{BrickRed}{$\times$} & 0.940 [0.874, 0.978] & 50.5 (3.3) & 1.4 (0.3) & 14.4 (3.3) & 1.5 (0.8) & 1.5 (0.8) & 0.032\\
\midrule
GoF-1 & \textcolor{BrickRed}{$\times$} & 0.940 [0.874, 0.978] & 52.9 (5.0) & 2.0 (0.2) & 14.0 (2.2) & 4.4 (0.8) & 4.6 (0.9) & 0.520\\
GoF-1.5 & \textcolor{BrickRed}{$\times$} & 0.930 [0.861, 0.971] & 39.2 (3.1) & 2.1 (0.2) & 11.6 (1.7) & 5.1 (0.8) & 5.3 (0.8) & 1.134\\
GoF-2 & \textcolor{BrickRed}{$\times$} & 0.880 [0.800, 0.936] & 21.8 (2.0) & 1.7 (0.2) & 8.4 (1.3) & 4.4 (0.6) & 4.5 (0.7) & 1.067\\
Simes & \textcolor{BrickRed}{$\times$} & 0.620 [0.517, 0.715] & 1.3 (0.2) & 1.0 (0.1) & 2.6 (0.3) & 2.5 (0.3) & 2.5 (0.3) & 1.117\\
\midrule
CountScan & \textcolor{BrickRed}{$\times$} & 0.950 [0.887, 0.984] & 40.7 (2.1) & \textbf{28.2} (1.2) & 0.0 (0.0) & 0.0 (0.0) & 0.0 (0.0) & 0.037\\
\bottomrule\end{tabular}
\caption{Simulation comparison at $(n,L,t)=(1024,128,1)$ over 100 replicates. Counts refer to the full document and the best 256-token window. Selected sizes use common $Y$, HMM, and CP rankings at FDP target $0.5$. CPU times cover preprocessing and 1,000 subset queries for $n=2048$, as described in Appendix~\ref{app:computational-evaluation}. Brackets give 95\% binomial intervals; parentheses give Monte Carlo standard errors. Bold marks column maxima.}\label{tab:sact-central}
\end{table}

Under Gumbel--max, SACT-$2$ and its CP variant have competitive detection power and generally certify more signals than unweighted scoring, $p$-value-based procedures, and CountScan (Figure~\ref{fig:sact-global}). Appending background text to a fixed watermarked passage leaves their counts nearly stable. After joint permutation of pivots, eligibility indicators, and signal labels and model refitting, both counts decline with length. SACT-$2$ then gives counts similar to GoF-1 (Figure~\ref{fig:sact-dilution}).

The gains from structural weighting are more pronounced for local inference. CountScan tends to certify more signals in short windows, while the SACT variants give larger bounds in longer windows (Figure~\ref{fig:sact-localization}). On the common HMM ranking at FDP target $0.5$, SACT-$2$ generally selects more tokens than unweighted scoring or the $q$-based score. BC and KJI--B can give larger selections at stricter targets (Figures~\ref{fig:sact-paths} and~\ref{fig:sact-app-paths-2048}).

Under SynthID-Text watermarking \citep{dathathri2024scalable}, both variants outperform CountScan in mean detection power and certified count across the 20 unpermuted signal settings. Their best 256-token windows have larger mean proportion bounds than under unweighted scoring or GoF-1 (Figures~\ref{fig:synthid-global} and~\ref{fig:synthid-localization}). Appendix~\ref{app:synthid} gives the implementation details.

\section{Methodological and Theoretical Foundations}\label{sec:theory}

We establish two theoretical results for SACT. First, under the stated conditional null assumptions, the bounds have finite-sample simultaneous coverage over all token sets, even when the working model for signal clustering is misspecified. Second, we derive an exact reference-family representation that provides an efficient shortcut for evaluating the full SACT-$2$ closed-testing bounds across repeated queries.

\subsection{Watermark mechanism and conditional null model}\label{subsec:mechanism-null}

We introduce the Gumbel--max verification statistic \citep{li2025framework,li2026robust}. Let $P_j=(P_{j,w})_{w\in\mathcal V}$ denote the next-token distribution given the prompt and the tokens preceding position $j$. Under the idealized randomness, the entries of $\zeta_j=(U_{j,w})_{w\in\mathcal V}$ are independent $\mathrm{Uniform}(0,1)$ variables conditional on $P_j$. The Gumbel--max decoder selects
\begin{equation}
\label{eq:wm-gumbel-decoder}
    W_j=\arg\max_{w:P_{j,w}>0}\{\log P_{j,w}-\log(-\log U_{j,w})\}=\arg\min_{w:P_{j,w}>0}\frac{-\log U_{j,w}}{P_{j,w}}.
\end{equation}
Averaging over the watermark randomness, this rule selects each token $w$ with probability $P_{j,w}$ and thus preserves the next-token distribution \citep{li2025framework}. The pivot $Y_j=U_{j,W_j}$ is the component of $\zeta_j$ indexed by the observed token $W_j$. If the observed token $W_j$ is independent of the verification vector $\zeta_j$, then $Y_j$ is uniform, with null distribution function $F_0(y)=y$ on $[0,1]$. For a token selected by the watermarked decoder in Equation~\eqref{eq:wm-gumbel-decoder}, the conditional distribution function given $P_j$ is
\begin{equation}\label{eq:wm-gumbel-alternative}
    F_{1,P_j}(y)=\sum_{w:P_{j,w}>0}P_{j,w}y^{1/P_{j,w}},\qquad 0\leq y\leq1.
\end{equation}
For example, if $P_j$ is uniform over $s$ tokens, then $Y_j\sim\mathrm{Beta}(s,1)$. If $P_j$ assigns all probability to one token, the pivot has the uniform null distribution even under watermarked generation. This variation in signal strength motivates the working model in Section~\ref{subsec:watermark-strength}.

Calibration across positions requires a joint model for the null pivots. When positions are filtered, we index the eligible positions in document order and condition throughout on their locations, which are used in fitting the working model. The following assumption concerns the pivots at these positions. Write $Y_I=(Y_j)_{j\in I}$ and $\Nset^c=[n]\setminus\Nset$.
\begin{assumption}[Conditional null pivots]\label{ass:null-pivots}
Conditional on the signal configuration $\xi_{1:n}$:
\begin{enumerate}
    \item[(i)] The null pivots are independent with common distribution $F_0$: $\mathcal L(Y_{\Nset}\mid\xi_{1:n})=F_0^{\otimes|\Nset|}$.

    \item[(ii)] The null and nonnull pivot vectors are independent: $Y_{\Nset}\perp\!\!\!\perp Y_{\Nset^c}\mid\xi_{1:n}$.
\end{enumerate}
\end{assumption}

The i.i.d.\ null model is a common working idealization in watermark detection \citep{li2025framework,li2026robust}. Assumption~\ref{ass:null-pivots}(i) applies this idealization to null positions conditional on the signal configuration, while part~(ii) requires that further conditioning on nonnull evidence leaves their joint distribution unchanged. This formulation is consistent with the latent-source model of \citet{li2026robust}, which separates ordinary-token sampling from watermark randomness. For a prespecified signal configuration, it is also consistent with the fixed-next-token-distribution idealization discussed by \citet{cai2025collision}, under which token-level pivots are treated as independent. Intuitively, it suffices that the reconstructed pseudorandom vectors $\zeta_j$ at null positions are mutually independent and independent of the completed document and its generation randomness. For example, suppose an author without access to the watermark key appends an independently written passage to watermarked text. Assume that the appended passage uses distinct seeding contexts not shared with the watermarked text. Under the idealization of independent randomness across contexts, the appended passage therefore has independent uniform pivots $Y_j$, independent of those in the watermarked passage.

\subsection{Sign-flip symmetry and calibration}\label{subsec:score-symmetry}

Null sign flips leave the fitted weights unchanged because they depend only on score magnitudes. Under Assumption~\ref{ass:null-pivots}, this yields the following symmetry.

\begin{property}[Sign-flip symmetry]\label{prop:sign-flip}
Suppose Assumption~\ref{ass:null-pivots} holds and $F_0$ is continuous. Conditional on any fixed signal indicators $\xi_{1:n}$, the scores in Equation~\eqref{eq:sign-flip-score} satisfy
\begin{equation}\label{eq:score-sign-flip}
    (X_1,\ldots,X_n)
    \overset{d}{=}
    (\eta_1X_1,\ldots,\eta_nX_n)
\end{equation}
for every fixed $\eta\in\{-1,1\}^n$ with $\eta_j=1$ for $j\notin\Nset$. Equivalently, conditional on $\xi_{1:n}$, all score magnitudes $(|X_j|)_{j\in[n]}$, and the nonnull scores $(X_j)_{j\notin\Nset}$, the null signs $(\operatorname{sign}(X_j))_{j\in\Nset}$ are independent and take values $+1$ and $-1$ with equal probability.
\end{property}

To see this, condition on $\xi_{1:n}$, the nonnull pivots, and the magnitudes $A_{1:n}$ used for fitting. These quantities determine the fitted weights, while the null signs remain independent and equally likely to be positive or negative. Since the weights are strictly positive, weighting preserves the signs. Reversing any subset of null signs leaves the conditional score distribution unchanged. Averaging over $A_{1:n}$ and the nonnull pivots gives Equation~\eqref{eq:score-sign-flip}.
The same calibration arguments apply to other mechanisms with conditionally sign-symmetric scores, including the centered binomial score for SynthID-Text defined in Appendix~\ref{app:synthid}.

The scan in Equation~\eqref{eq:sign-scan-statistic} has the same form as the CUSUM sign statistic of \citet{arias2017distribution}. Property~\ref{prop:sign-flip} gives its null reference distribution for the structure-adaptive scores. Let $e_1,\ldots,e_m$ be independent Rademacher variables, each taking values $+1$ and $-1$ with equal probability. For $m\geq1$ and $0<\alpha<1/2$, define
\begin{equation}\label{eq:sign-scan-null}
    T_m^0=\max_{1\leq k\leq m}
          \frac{\sum_{\ell=1}^k e_\ell}{\sqrt{k}},
    \qquad
    c_{\alpha,m}
    =\min\left\{c\in\operatorname{supp}(T_m^0):
          \Pr(T_m^0\geq c)\leq\alpha\right\},
\end{equation}
with $c_{\alpha,m}=+\infty$ if the displayed set is empty. Write $c_m=c_{\alpha,m}$ and set $c_0=+\infty$. For nonempty $I$, the local test is $\phi_I^{\mathrm{scan}}=\ind\{T_I\geq c_{\alpha,|I|}\}$; set $\phi_\varnothing^{\mathrm{scan}}=0$.

Under $H_I$, every position in $I$ is null. Conditional on $\xi_{1:n}$ and the score magnitudes, the ordering is fixed and the ordered signs in $I$ are independent Rademacher variables. Thus, $T_I$ has the conditional distribution of $T_{|I|}^0$, and the local test rejects with conditional probability at most $\alpha$.

\subsection{Simultaneous validity}
\label{sec:prelim}

Write $d_\alpha^{\mathrm{CT}}(R)$ and $\underline{\TDP}^{\mathrm{CT}}_\alpha(R)$ for the closed-testing count and proportion bounds in Section~\ref{subsec:sign-scan-test}, with $[n]$ denoting the testable positions. The following theorem establishes their simultaneous validity.

\begin{theorem}[Simultaneous validity of SACT]\label{thm:sact-validity}
Suppose the scores have the sign-flip symmetry stated in Property~\ref{prop:sign-flip}, and let $0<\alpha<1/2$. Construct the scores in Equation~\eqref{eq:sign-flip-score} with strictly positive weights depending only on $A_{1:n}$, and resolve ties in score magnitudes by a fixed rule independent of the signs. For the sign-scan local tests in Section~\ref{subsec:score-symmetry}, calibrated using the exact critical values in Equation~\eqref{eq:sign-scan-null}, the closed-testing bounds satisfy
\begin{equation*}
    \Pr\big\{\underline{\TDP}^{\mathrm{CT}}_\alpha(R)\leq\TDP(R)\text{ for every nonempty }R\subseteq[n]\big\}\geq1-\alpha.
\end{equation*}
The guarantee holds in finite samples and does not require a correctly specified working model.
\end{theorem}

The argument follows \citet{goeman2011exploratory}. With probability at least $1-\alpha$, the local test does not reject the true null set $\Nset$. On this event, $\Nset$ is a candidate in the minimization in Equation~\eqref{eq:database-count-bound}, so $d_\alpha^{\mathrm{CT}}(R)\leq |R\setminus\Nset|=d(R)$ for every $R$ simultaneously.

\subsection{An exact reference-family representation}
\label{subsec:reference-family}

A reference family interpolates jointly valid signal-count lower bounds to obtain proportion bounds for arbitrary sets \citep{blanchard2020reference}. Let $\mathfrak R=\{(S_k,\ell_k):k\in\mathcal K\}$ consist of sets $S_k\subseteq[n]$ and integer lower bounds $0\leq\ell_k\leq|S_k|$ satisfying
\begin{equation}\label{eq:reference-joint}
    \Pr\big\{d(S_k)\geq\ell_k\text{ for all }k\in\mathcal K\big\}\geq1-\alpha.
\end{equation}
On this event, at most $|S_k\setminus R|$ of the signals certified in $S_k$ can lie outside $R$. For nonempty $R$, taking the strongest remaining bound gives
$$
    \underline{\TDP}^{\mathrm{RF}}_\alpha(R)
    =\frac{1}{|R|}\max_{k\in\mathcal K}\bigl(\ell_k-|S_k\setminus R|\bigr)_+,
    \qquad
    d_\alpha^{\mathrm{RF}}(R)=|R|\underline{\TDP}^{\mathrm{RF}}_\alpha(R).
$$
These bounds hold simultaneously over all $R$. We now construct an SACT-$2$ reference family whose interpolation recovers the full closed-testing bounds.

Suppose $0<\alpha<1/2$ and the fitted scores are nonzero with pairwise distinct absolute values. Use the exact critical values in Equation~\eqref{eq:sign-scan-null}, and let $\pi$ order the positions by decreasing $|X_j|$. For $k\in\mathcal K=\{0,\ldots,n\}$, define $S_k=\{\pi(\ell):1\leq\ell\leq k,\varepsilon_{\pi(\ell)}=1\}$, $P_k=|S_k|$, and $Q_k=k-P_k$. Thus, $S_k$ contains the positive-score positions among the first $k$ ranks, and $Q_k$ counts the negative scores among those ranks. The sets $S_k$ are nested.

Let $m_0=|\Nset|$ be the unknown number of null positions. For a candidate null-set size $r$ with finite critical value $c_r$, the SACT-$2$ scan gives the signal-count lower bound
\begin{equation}\label{eq:scan-prefix-boundary}
    \ell_k(r)=\left[P_k-\min\left\{r,
       \left\lceil\frac{r+c_r\sqrt r}{2}\right\rceil-1,
       \left\lceil Q_k+\frac{c_r^2}{2}
          +c_r\sqrt{2Q_k+\frac{c_r^2}{4}}\right\rceil-1\right\}\right]_+.
\end{equation}
Intuitively, nonrejection of the scan limits how many positive scores in $S_k$ can be explained by a null set of size $r$. Any excess of $P_k$ over this limit gives a lower bound on the retained-signal count. Set $\ell_k(0)=P_k$ and $\ell_k(r)=(P_k-r)_+$ when $c_r=+\infty$.

If the scan does not reject the true null set, then $d(S_k)\geq\ell_k(m_0)$ for every $k$. Since the total signal count is $n-m_0$, this event also implies $m_0+\ell_k(m_0)\leq n$. We therefore take the largest candidate size consistent with all ranks,
\begin{equation}\label{eq:global-feasible-size}
    \widehat m=\max\left\{r\in\{0,\ldots,n\}:
       r+\ell_k(r)\leq n\text{ for all }0\leq k\leq n\right\},
\end{equation}
and store the corresponding reference bounds
\begin{equation}\label{eq:stored-scan-boundary}
    \ell_k=\ell_k(\widehat m),\qquad 0\leq k\leq n.
\end{equation}
By Lemma~\ref{lem:scan-boundary-nesting}, $\ell_k(r)$ is nonincreasing in $r$. Since $m_0\leq\widehat m$ on the nonrejection event, $d(S_k)\geq\ell_k(m_0)\geq\ell_k$ for all $k$, which establishes the joint guarantee in Equation~\eqref{eq:reference-joint}. The following theorem shows that interpolation through this reference family yields exactly the same lower bounds as the full closed-testing procedure for every selected set, without additional conservatism.

\begin{theorem}[Reference-family representation of the SACT bounds]\label{thm:computational-bound}
Suppose $0<\alpha<1/2$ and the fitted scores are nonzero with pairwise distinct absolute values. Use the full sign scan and exact critical values in Equation~\eqref{eq:sign-scan-null}. Then the reference family with lower bounds given by Equation~\eqref{eq:stored-scan-boundary} satisfies $d_\alpha^{\mathrm{RF}}(R)=d_\alpha^{\mathrm{CT}}(R)$ for every $R\subseteq[n]$.
\end{theorem}
The reference-family representation in Theorem~\ref{thm:computational-bound} recovers the full closed-testing bounds without enumerating all intersection hypotheses. During an interactive review, readers may query the same document many times as they compare passages or revise their selections. Each query reuses the reference sets and lower bounds computed once for that document.

Specifically, after fitting and calibration, constructing the reference family for $n$ tokens takes $O(n^2)$ time, and evaluating Equation~\eqref{eq:database-interpolation} for each query takes $O(n)$ time. The total cost for $M$ sets is therefore $O(n^2+Mn)$. Exact equivalence also preserves admissibility whenever the underlying closed-testing procedure is admissible \citep{goeman2021admissible}. Appendix~\ref{app:proofs} gives the proofs and algorithms, including extensions to power-divergence and BC scans under their calibration and monotonicity conditions.

\section{Conclusion}
\label{sec:con}

We developed simultaneous lower confidence bounds on the proportion of tokens retaining a watermark signal in any nonempty set of token positions. SACT exploits sequential signal clustering. Readers can examine the text and watermark evidence and refine their selections while retaining the same coverage guarantee. Under the stated null conditions, this guarantee holds in finite samples even if the working model is misspecified. Computing the reference family once per document enables reviewers to recover the full closed-testing bound efficiently for each selection. In simulations under Gumbel--max and SynthID-Text watermarking, structural weighting generally improved the passage-level bounds. The database-recovery review showed that assessing related passages together can sharpen the lower bound on their combined count of retained signals.

An important future direction is to extend the null model to account for dependence arising from repeated seeding contexts and editing. This includes characterizing when conditional sign symmetry is preserved, developing calibration methods for dependent null scores, and quantifying coverage error under departures from symmetry. Further theoretical work could relate signal strength, cluster length, and document length to detection power and the tightness of simultaneous bounds, clarifying when structure adaptation is most effective. Systematic studies of substitutions, insertions, deletions, and paraphrasing would complement this analysis by assessing how effectively SACT combines the watermark evidence that survives editing.

\bibliography{WM}
\label{page:main-end}

\clearpage
\begin{center}
{\Large\bfseries Supplementary Material for \Title}
\end{center}

\appendix
\numberwithin{equation}{section}
\counterwithin{figure}{section}
\counterwithin{table}{section}

Appendix~\ref{app:proofs} establishes conditions under which the proposed algorithms recover the full closed-testing bounds. Appendix~\ref{app:technical-details} describes the implementation of the methods used in the case study and simulations. Appendix~\ref{app:sact-experiments} presents the simulation design and results on simultaneous coverage, detection power, and the effect of signal clustering on the bounds.

\section{Proofs}\label{app:proofs}

We first prove that the reference-family bound for SACT-$2$ equals the full closed-testing bound and extend the argument to other sign scans. We then give a separate algorithm for symmetric monotone local tests, which we use to compute the GoF and Simes bounds.

\subsection{Reference-family representation of SACT-2}\label{app:proofs-computation}

For a nonempty query set $R$, a reference family satisfying the joint guarantee in Equation~\eqref{eq:reference-joint} gives the proportion and count lower bounds
\begin{equation}\label{eq:reference-interpolation}
    \underline{\TDP}^{\mathrm{RF}}_\alpha(R)=\frac{1}{|R|}\max_{k\in\mathcal K}\bigl(\ell_k-|S_k\setminus R|\bigr)_+,\qquad d_\alpha^{\mathrm{RF}}(R)=|R|\underline{\TDP}^{\mathrm{RF}}_\alpha(R).
\end{equation}
Set $d_\alpha^{\mathrm{RF}}(\varnothing)=0$. In the proofs and algorithms below, $[n]$ indexes the eligible positions. For a selection $R$ in the original document, apply the algorithm to its eligible positions $R_e=R\cap\mathcal E$. If $R_e$ is empty, the certified count is zero. Otherwise, a proportion bound $L$ for $R_e$ gives a count bound of $|R_e|L$ for $R$. Divide this count by the full selection size $|R|$ to obtain the proportion bound for any nonempty $R$.

\paragraph*{Construction of signal-count lower bounds.}
We use the notation of Section~\ref{subsec:reference-family}, relabeling positions by decreasing score magnitude so that $S_k=\{i\leq k:\varepsilon_i=1\}$. Write $P=P_n$ and $H=Q_n$. Fix a rank $k$ and let $q=Q_k$ be the number of negative scores among the first $k$ positions. Suppose a candidate null set of size $r$ contains $v$ of the positive scores among these positions. The candidate has $r-v$ places left, so it can include at most $u_r(v,q)=\min\{q,r-v\}$ of the $q$ negative scores. For a given $v$, including that many negatives gives the smallest standardized sign sum. Since $v+u_r(v,q)=\min\{r,v+q\}$, the largest $v$ allowed by this constraint is
\begin{equation}\label{eq:scan-count-envelope}
    G_r(q)=
    \max\left\{v\in\{0,\ldots,r\}:
       v<\frac{\min\{r,v+q\}+c_r\sqrt{\min\{r,v+q\}}}{2}\right\}.
\end{equation}
Include $v=0$ by convention, set $G_0(q)=0$, and take $G_r(q)=r$ when $c_r=+\infty$. When $v+u_r(v,q)>0$, the same constraint can be written as $(v-u_r(v,q))/\sqrt{v+u_r(v,q)}<c_r$. Subtracting the largest feasible count $G_r(Q_k)$ from the observed positive count $P_k$ gives $\ell_k(r)=(P_k-G_r(Q_k))_+$. The stored lower bound is therefore $\ell_k=(P_k-G_{\widehat m}(Q_k))_+$. Every finite $c_r$ is positive because $T_r^0\geq1$ whenever the first sign is positive, an event with probability $1/2>\alpha$.

For finite $c>0$, write $L_c(q)=\lceil q+c^2/2+c\sqrt{2q+c^2/4}\rceil-1$. Then
\begin{equation}\label{eq:scan-boundary-closed}
    G_r(q)=\min\left\{r,L_{c_r}(q),
        \left\lceil\frac{r+c_r\sqrt r}{2}\right\rceil-1\right\}.
\end{equation}
The second term uses the fact that only $q$ negative scores are available. The third uses the fact that the candidate has only $r$ positions. For an integer $v$, the strict bound $v<a$ is equivalent to $v\leq\lceil a\rceil-1$. The following lemma shows that the feasible positive counts form an initial interval. We use this property below to construct a nonrejected set attaining the closed-testing bound.

\begin{lemma}[Boundary calculation and initial intervals]\label{lem:scan-boundary-calculation}
For finite $c_r>0$, Equations~\eqref{eq:scan-boundary-closed} and~\eqref{eq:scan-count-envelope} give the same value of $G_r(q)$. For each $r,q$, the feasible integers $v$ in Equation~\eqref{eq:scan-count-envelope} form an initial interval starting at zero.
\end{lemma}
\begin{proof}
For $v+u>0$, the ratio $(v-u)/\sqrt{v+u}$ increases with $v$ and decreases with $u$. As $v$ increases, $u_r(v,q)$ cannot increase. Once a value of $v$ violates the constraint, every larger value does too. This proves the initial-interval property, with the empty state included by convention.

To obtain the explicit boundary, first allow all $q$ negatives. For $v>q$, solving $v-q<c\sqrt{v+q}$ gives $v<q+c^2/2+c\sqrt{2q+c^2/4}$; for $v\le q$, the inequality holds automatically outside the empty state. Thus the largest feasible integer is $L_c(q)$. For $r>0$ and $v+q>0$, imposing the second limit of $r-v$ negatives gives
\[
 \frac{v-\min(q,r-v)}{\sqrt{v+\min(q,r-v)}}
 =\max\left\{\frac{v-q}{\sqrt{v+q}},\frac{2v-r}{\sqrt r}\right\}.
\]
Hence the constraint holds exactly when both terms on the right are below $c_r$, or equivalently,
\[
 v\le L_{c_r}(q),\qquad
 v<\frac{r+c_r\sqrt r}{2},
\]
together with $v\le r$, proving Equation~\eqref{eq:scan-boundary-closed}. The empty-state convention agrees with these nonnegative bounds.
\end{proof}

\begin{lemma}[Monotonicity of the signal-count lower bounds]\label{lem:scan-boundary-nesting}
With the exact critical values in Equation~\eqref{eq:sign-scan-null}, $\ell_k(r)\geq\ell_k(r+1)$ for all $0\leq k\leq n$ and $0\leq r<n$.
\end{lemma}

\begin{proof}[Proof of Lemma~\ref{lem:scan-boundary-nesting}]
Since $\ell_k(r)=(P_k-G_r(Q_k))_+$, it suffices to show $G_r(q)\le G_{r+1}(q)$ for every $q$. Fix a feasible $v$ at size $r\ge1$ and put $u=\min(q,r-v)$. The empty state remains feasible. Otherwise, set $s=(v-u)/\sqrt{v+u}$. If $s\le0$, increasing $r$ allows at least as many negatives and cannot increase the ratio, so feasibility follows from positivity of every finite cutoff.

For $s>0$, consider the length-$r$ sequence of $u$ negatives, then $v$ positives, then $r-u-v$ negatives. Its scan maximum is $s$: the positive block increases the standardized sum to $s$, and the final negatives decrease it. Appending one negative preserves this maximum. Consequently, $s$ is an attainable value for both $T_r^0$ and $T_{r+1}^0$.

Because $s<c_r$ is attainable, the canonical definition of $c_r$ implies $\Pr(T_r^0\ge s)>\alpha$, including when $c_r=+\infty$. Coupling the two null walks through their first $r$ signs gives $T_{r+1}^0\ge T_r^0$, and therefore
\[
    s<c_r\ \Longrightarrow\
    \Pr(T_{r+1}^0\ge s)\ge\Pr(T_r^0\ge s)>\alpha
    \ \Longrightarrow\ s<c_{r+1}.
\]
The last implication follows from the definition of the cutoff because a walk of length $r+1$ can attain $s$. The ratio for $v$ at size $r+1$ is no larger than $s$, so $v$ remains feasible. Thus $G_r(q)\le G_{r+1}(q)$ and the signal-count lower bounds are nonincreasing; the case $r=0$ is immediate.
\end{proof}

\paragraph*{Equivalence with full closed testing.}
\begin{proof}[Proof of Theorem~\ref{thm:computational-bound}]
We first show that the complement of every locally nonrejected set satisfies the stored signal-count lower bounds, giving one inequality between the two procedures. We then construct a nonrejected set that attains the reference-family bound for each query, proving the reverse inequality.

\textit{Step 1: Every nonrejected candidate satisfies the reference constraints.}
Let $\mathcal A=\{I\subseteq[n]:T_I<c_{|I|}\}$.
Fix $I\in\mathcal A$ of size $r$. Among the first $k$ ranked positions, let
\[
 v=|I\cap S_k|,\qquad u=|\{i\in I:i\le k,\ \varepsilon_i=-1\}|.
\]
In the ordering of $I$, these are the first $v+u$ scores. If $v+u>0$, their standardized sign sum is $(v-u)/\sqrt{v+u}$. It is below $c_r$ because $I$ was not rejected. Replacing $u$ by $\min(Q_k,r-v)$ leaves the sum below $c_r$, so $v\le G_r(Q_k)$. This bound also holds when $v=u=0$. Consequently,
\[
 \ell_k(r)=(P_k-G_r(Q_k))_+\le P_k-v=|S_k\setminus I|\le n-r.
\]
Thus $r+\ell_k(r)\le n$ for every $k$, so $r\le\widehat m$. Lemma~\ref{lem:scan-boundary-nesting} gives $|S_k\setminus I|\ge\ell_k(r)\ge\ell_k$. For every query $R$ and every $k$,
\[
 |R\setminus I|\ge |S_k\setminus I|-|S_k\setminus R|\ge\ell_k-|S_k\setminus R|.
\]
Minimizing over nonrejected $I$ proves
\begin{equation}\label{eq:rf-closure-lower}
 d_\alpha^{\mathrm{CT}}(R)\ge d_\alpha^{\mathrm{RF}}(R).
\end{equation}

\textit{Step 2: Determine how many positive scores to select at the largest feasible size.}
Write $r_*=\widehat m$ and $p_*=r_*-H$. The set of all $H$ positions with negative scores is not locally rejected, so Step 1 gives $r_*\ge H$ and $p_*\ge0$. Let $S_+=\{i:\varepsilon_i=1\}$. For $A\subseteq S_+$ define
\begin{equation}\label{eq:rf-positive-quota}
 K(A)=\min\left\{|A|,\min_{0\le k\le n}
 \bigl(|A\setminus S_k|+P_k-\ell_k\bigr)\right\}.
\end{equation}
Feasibility of $r_*$ gives
\[
 p_*\le P-\ell_k\quad\text{for every }k,
\]
so $K(S_+)\ge p_*$. If $r_*<n$ and $K(S_+)>p_*$, integrality gives $P-\ell_k\ge p_*+1$ for every $k$. Monotonicity implies
\[
 r_*+1+\ell_k(r_*+1)\le r_*+1+\ell_k\le n,
\]
so $r_*+1$ would be feasible, a contradiction. If $r_*=n$, then $K(S_+)\le P=p_*$. In either case $K(S_+)=p_*$. The formula for $K$ is monotone under inclusion, and hence $0\le K(A)\le p_*$ for every $A\subseteq S_+$.

\textit{Step 3: Construct a nonrejected set for each query.}
Fix $R$ and set $A=R\cap S_+$ and $K=K(A)$. Include all $H$ positions with negative scores. Among the positions in $A$, take the $K$ with the smallest score magnitudes. Fill the remaining places with the smallest positive scores still available, until the set contains $p_*$ positions with positive scores. This is possible because $K\le p_*\le P$. Let $I_R$ denote the resulting set of size $r_*$.

Among the first $k$ ranked positions, the number of positive scores selected for $I_R$ is
\begin{equation}\label{eq:rf-witness-prefix}
 |I_R\cap S_k|=
 \max\left\{\positivepart{K-|A\setminus S_k|},\
                 \positivepart{p_*-(P-P_k)}\right\}.
\end{equation}
Beyond rank $k$, there are $|A\setminus S_k|$ positions from $A$ and $P-P_k$ positive scores in total. To select $K$ positions from $A$, we therefore need at least the first term from ranks $1$ through $k$. To select $p_*$ positive scores in total, we need at least the second term from those ranks. Choosing the smallest scores first gives exactly the larger of these two counts.

The first term in Equation~\eqref{eq:rf-witness-prefix} is at most $P_k-\ell_k$ by Equation~\eqref{eq:rf-positive-quota}; the second satisfies the same inequality by feasibility of $r_*$. Hence $v=|I_R\cap S_k|\le P_k-\ell_k=\min\{P_k,G_{r_*}(Q_k)\}$. Since $v\le p_*$,
\[
 r_*-v\ge H\ge Q_k,\qquad
 u_{r_*}(v,Q_k)=Q_k.
\]
Among the first $k$ positions, $I_R$ contains $v$ positive scores and all $Q_k$ negative scores. If $v+Q_k>0$, the scan on $I_R$ at rank $v+Q_k$ therefore has value $(v-Q_k)/\sqrt{v+Q_k}$. This is below $c_{r_*}$ by Lemma~\ref{lem:scan-boundary-calculation}, since $v\le G_{r_*}(Q_k)$. As $k$ runs from $0$ to $n$, $v+Q_k$ goes up by one each time we reach a position in $I_R$. This accounts for every rank in the scan, so $I_R\in\mathcal A$. If $r_*=0$, the empty set belongs to $\mathcal A$ by convention.

\textit{Step 4: Show that the constructed set attains the bound.}
The set $I_R$ contains every negative-score position in $R$ and at least $K$ of its positive-score positions. Hence
\[
 d_\alpha^{\mathrm{CT}}(R)\le |R|-|I_R\cap R|\le |A|-K.
\]
By Equation~\eqref{eq:rf-positive-quota} and $A=R\cap S_+$, $|A|-K=\max_k(\ell_k-|S_k\setminus R|)_+=d_\alpha^{\mathrm{RF}}(R)$. Combining this with Equation~\eqref{eq:rf-closure-lower} proves equality with full closure and shows that $I_R$ attains the bound. By Step 1, every nonrejected set has at most $r_*$ elements. The set constructed in Step 3 has exactly $r_*$, so $r_*=\max_{I\in\mathcal A}|I|$. Finally, monotonicity gives $\min_{r\le\widehat m}\ell_k(r)=\ell_k(\widehat m)$.
\end{proof}

A scan value equal to the cutoff leads to rejection, as reflected in the strict inequality in Equation~\eqref{eq:scan-count-envelope}. If $c_r=a/\sqrt b>0$ and $v>u$, the scan value at this rank is below the cutoff exactly when $b(v-u)^2<a^2(v+u)$. This comparison can be made with integer or certified arithmetic to avoid rounding errors at the cutoff. The complexity statements count arithmetic operations.

\paragraph*{Preprocessing and repeated queries.}
Algorithm~\ref{alg:sact-bound} implements the exact representation. It computes the ordering and reference-family lower bounds once, then reuses them for each selected token set.

\begin{algorithm}[!ht]
\caption{SACT-$2$: preprocessing and repeated queries}\label{alg:sact-bound}
\spacingset{1.1}
\small
\begin{algorithmic}
\Require Nonzero scores with distinct magnitudes; exact cutoffs $c_0,\ldots,c_n$.
\Procedure{Preprocess}{$X_{1:n},c_{0:n}$}
    \State Sort the scores by decreasing magnitude and compute the cumulative counts $P_k$ and $Q_k$ of positive and negative scores.
    \State Set $\widehat m\gets0$.
    \For{$r=1,\ldots,n$}
        \State Evaluate $\ell_k(r)$ for all $k$ by Equation~\eqref{eq:scan-prefix-boundary} and its conventions.
        \If{$r+\ell_k(r)\leq n$ for every $k=0,\ldots,n$}
            \State Set $\widehat m\gets r$.
        \EndIf
    \EndFor
    \State Set $\ell_k\gets\ell_k(\widehat m)$ for $k=0,\ldots,n$.
    \State \Return $(\pi,\varepsilon,\ell)$, with $\varepsilon_j=\operatorname{sign}(X_j)$.
\EndProcedure
\Procedure{Query}{$R,\pi,\varepsilon,\ell$}
    \State Form a membership mask for $R$; set $a\gets|R|$, $C\gets0$, and $d\gets0$.
    \For{$k=1,\ldots,n$}
        \If{$\varepsilon_{\pi(k)}=1$ and $\pi(k)\notin R$}
            \State Set $C\gets C+1$.
        \EndIf
        \State Set $d\gets\max\{d,\ell_k-C\}$.
    \EndFor
    \State \Return $(d,d/(a\vee1))$.
\EndProcedure
\end{algorithmic}
\end{algorithm}

\FloatBarrier

\subsection{Extensions to other sign scans}\label{app:scan-extensions}

We extend the reference-family construction to power-divergence scans and the Barber--Cand\`es scan.

\subsubsection{Power-divergence scans}\label{app:divergence-reference}

For the power-divergence scans in Equation~\eqref{eq:power-divergence-sign-scan}, we apply the construction in Section~\ref{subsec:reference-family} with each scan's own signal-count lower bounds. For a candidate of size $r$, consider the first $k$ positions in the magnitude ranking. Set $u=\min\{Q_k,r-v\}$ and use binary search to find the largest positive count $0\le v\le r$ whose contribution $(v+u)d_s\{v/(v+u),1/2\}$ is below the cutoff for testing a set of size $r$, taking the contribution as zero when $v\le u$. This search works because the contribution is nondecreasing in $v$. While $u=Q_k$, adding a positive increases both the total count and the positive fraction. Once $u=r-v$, the total stays at $r$ and the positive fraction continues to increase. If the largest feasible count is $v_{\max}$, the signal-count lower bound is $\ell_k(r)=(P_k-v_{\max})_+$. If $\ell_k(r)$ is nonincreasing in $r$ for each rank $k$, the argument in Appendix~\ref{app:proofs-computation} shows that the reference family gives the same bounds as full closed testing.

\subsubsection{Barber--Cand\`es procedure}
\label{app:bc-reference}

Let $c_r^{\mathrm{BC}}$ be the smallest point in the support of the length-$r$ fair-sign BC scan with inclusive upper-tail probability at most $\alpha$, or $+\infty$ if no such point exists. Set $c_0^{\mathrm{BC}}=+\infty$. Define
\[
    G_r^{\mathrm{BC}}(q)=
    \max\left\{v\in\{0,\ldots,r\}:
       \frac{v}{1+\min(q,r-v)}<c_r^{\mathrm{BC}}\right\},
\]
with $G_0^{\mathrm{BC}}(q)=0$ and $G_r^{\mathrm{BC}}(q)=r$ for an infinite cutoff. For a finite cutoff $c=c_r^{\mathrm{BC}}$,
\begin{equation}\label{eq:bc-count-envelope}
    G_r^{\mathrm{BC}}(q)=
    \min\left\{r,\left\lceil c(1+q)\right\rceil-1,
       \left\lceil\frac{c(r+1)}{1+c}\right\rceil-1\right\}.
\end{equation}

\begin{proposition}[BC representation and complexity]\label{prop:bc-reference}
For nonzero scores with distinct magnitudes and $0<\alpha<1/2$, using $\ell_k^{\mathrm{BC}}(r)=(P_k-G_r^{\mathrm{BC}}(Q_k))_+$ in place of $\ell_k(r)$ in Equations~\eqref{eq:global-feasible-size}--\eqref{eq:stored-scan-boundary} gives a reference family whose interpolation equals full closure with the exact BC local tests. These signal-count lower bounds are nonincreasing in $r$. Constructing the reference family takes $O(n^2)$ time, and evaluating its bound for one query set takes $O(n)$ time.
\end{proposition}

\begin{proof}
Feasible positive counts form an initial interval because $v/(1+u)$ increases with $v$ and decreases with $u$. The identity
\[
    \frac{v}{1+\min(q,r-v)}
    =\max\left\{\frac{v}{1+q},\frac{v}{1+r-v}\right\}
\]
gives the constraints $v<c(1+q)$ and $v<c(r+1)/(1+c)$, proving Equation~\eqref{eq:bc-count-envelope}.

For a feasible $v>0$, put $u=\min(q,r-v)$. The sign sequence of $u$ negatives, $v$ positives, and $r-u-v$ negatives has BC maximum $s=v/(1+u)$, unchanged by appending a negative. The calibration argument in Lemma~\ref{lem:scan-boundary-nesting} gives $s<c_{r+1}^{\mathrm{BC}}$. At size $r+1$ we can include at least as many negatives, so $v$ remains feasible. The count $v=0$ is feasible by convention.

These arguments provide the two properties used in the proof of Theorem~\ref{thm:computational-bound}: feasible counts form initial intervals, and the signal-count lower bounds are nonincreasing in candidate size. We can therefore construct the same nonrejected set using $G_r^{\mathrm{BC}}$ in place of $G_r$. This proves equality with full closure. Finally, Equation~\eqref{eq:bc-count-envelope} evaluates each lower bound in constant time, giving the same preprocessing and query costs.
\end{proof}

\subsection{Closed-testing bounds for symmetric monotone local tests}
\label{app:querywise-shortcut}
\label{subsec:closed-computation}

For the GoF and Simes tests in Section~\ref{sec:numerical}, we use Algorithm~\ref{alg:sact-querywise} to compute the closed-testing bound for each query without constructing a reference family. Both local tests are permutation invariant and coordinatewise nondecreasing in the evidence scores $X_j=-p_j$, since decreasing a $p$-value preserves any rejection. The algorithm also applies to the sign-based tests under the conditions given below. For the tests considered here, it takes $O(n^2)$ time per query. For repeated queries with SACT-$2$, Algorithm~\ref{alg:sact-bound} is more efficient because it constructs the reference family once in $O(n^2)$ time and then evaluates each query in $O(n)$ time.

\begin{algorithm}[!ht]
\caption{Exact closed-testing bounds for a single query}\label{alg:sact-querywise}
\spacingset{1.1}
\small
\algtext*{EndIf}
\algtext*{EndWhile}
\begin{algorithmic}
\Require Fixed evidence scores $X_{1:n}$; nonempty $R$; local tests $\phi_I=h_{|I|}(X_I)$ with symmetric, coordinatewise monotone $h_m$.
\Ensure Closed-testing lower confidence bound $\underline{\TDP}^{\mathrm{CT}}_\alpha(R)$.
\State Set $r\gets |R|$, $m\gets n$, and $b\gets0$.
\State Sort $[n]$ as $(w_1,\ldots,w_n)$ so that $X_{w_1}\leq\cdots\leq X_{w_n}$.
\State Let $(u_1,\ldots,u_r)$ be the subsequence of $(w_1,\ldots,w_n)$ belonging to $R$.
\While{$b<\min(r,m)$}
    \State Set $s\gets b+1$ and $L_{m,s}\gets\{u_1,\ldots,u_s\}$.
    \State Add to $L_{m,s}$ the $m-s$ smallest-scoring positions in $[n]\setminus L_{m,s}$.
    \If{$\phi_{L_{m,s}}=1$}
        \State $m\gets m-1$
    \Else
        \State $b\gets |L_{m,s}\cap R|$
    \EndIf
\EndWhile
\State \Return $1-b/r$
\end{algorithmic}
\end{algorithm}

Algorithm~\ref{alg:sact-querywise} searches for the largest overlap between the query $R$ and an intersection that is not locally rejected. To organize this search, fix a nonempty $R$ with $r=|R|$ and define
\[
    \mathcal C_{m,s}(R)=\{I\subseteq[n]:|I|=m,\ |I\cap R|\ge s\},
    \qquad 1\le s\le\min(m,r).
\]
Construct $L_{m,s}$ by first taking the $s$ smallest evidence scores in $R$, then adding the $m-s$ smallest scores among all remaining positions. Some of the added positions may also belong to $R$, so the final set can overlap $R$ in more than $s$ positions. At each rank, the score in $L_{m,s}$ is the smallest possible among sets in this family. Testing $L_{m,s}$ therefore suffices to determine whether the family contains a nonrejected set. This is the symmetry-and-monotonicity principle underlying FACT \citep{dobriban2020fact}.

\begin{lemma}[Symmetry and monotonicity]\label{lem:sign-scan-monotonicity}
For fixed size $m$ and cutoff, the sign-scan, power-divergence, and BC rules based on Equations~\eqref{eq:sign-scan-statistic},~\eqref{eq:power-divergence-sign-scan}, and~\eqref{eq:bc-sign-scan} are permutation invariant and coordinatewise nondecreasing on nonzero score vectors with pairwise distinct magnitudes.
\end{lemma}
\begin{proof}
Fix the size $m$ and its cutoff. Distinct magnitudes determine a unique ordered sign sequence, regardless of coordinate labels, giving permutation invariance.

Now increase one signed score while holding the others fixed. If it is negative, its magnitude decreases and it moves later in the ordering; if positive, it moves earlier. Exchanges with a score of the same sign do not change the sign sequence. An exchange of opposite signs replaces $(-1,+1)$ by $(+1,-1)$, increasing the cumulative positive count at the first of the two ranks and leaving the counts at all other ranks unchanged. If the score crosses zero, it changes from negative to positive at the last rank. Thus, for every $k$, the number of positive signs among the first $k$ ranks cannot decrease.

At rank $k$, the contribution to the scan statistic is nondecreasing in the number of positive signs among the first $k$ scores. This holds directly for the standardized sign sum and BC ratio, and for the power-divergence contribution above $k/2$, with zero contribution below that threshold. Taking the maximum over $k$ and applying the fixed cutoff preserves monotonicity.
\end{proof}

\begin{theorem}[Exact computation for a single query]\label{thm:querywise-bound}
Fix evidence scores $X_{1:n}$ and local rules $\phi_I=h_{|I|}(X_I)$ that are permutation invariant and coordinatewise nondecreasing. Algorithm~\ref{alg:sact-querywise} returns the proportion bound $d_\alpha^{\mathrm{CT}}(R)/|R|$ corresponding to Equation~\eqref{eq:database-count-bound} using at most $n$ local-test evaluations. For the sign, power-divergence, BC, GoF, and Simes tests, its cost is $O(n^2)$ per query after fitting and calibration.
\end{theorem}
\begin{proof}
We first show that the ordered scores of $L_{m,s}$ are coordinatewise smallest among the sets in $\mathcal C_{m,s}(R)$. For a threshold $t$, write $a(t)=|\{j\in R:X_j\le t\}|$ and $c(t)=|\{j\notin R:X_j\le t\}|$. A member of $\mathcal C_{m,s}(R)$ can contain at most $a(t)+c(t)$ scores no greater than $t$. Moreover, it must take at least $[s-a(t)]_+$ query positions whose scores exceed $t$, leaving at most $m-[s-a(t)]_+$ positions at or below $t$. The construction of $L_{m,s}$ attains the smaller of these two limits for every $t$, so every $I\in\mathcal C_{m,s}(R)$ satisfies
\[
    |\{j\in I:X_j\le t\}|
    \le\min\{a(t)+c(t),\ m-[s-a(t)]_+\}
    =|\{j\in L_{m,s}:X_j\le t\}|.
\]
Consequently, the ordered scores of $L_{m,s}$ are coordinatewise no greater than those of $I$. Symmetry and monotonicity imply $\phi_{L_{m,s}}\le\phi_I$. Since $L_{m,s}$ itself belongs to the family,
\begin{equation}\label{eq:least-favourable-family}
    \exists I\in\mathcal C_{m,s}(R):\phi_I=0
    \quad\Longleftrightarrow\quad\phi_{L_{m,s}}=0.
\end{equation}

Next, let $b^*=\max_{\phi_I=0}|I\cap R|$ be the overlap sought by the algorithm. At each iteration, $b$ is an overlap already attained by a nonrejected set. Any nonrejected set with a larger overlap must have size at most $m$. Initially $m=n$, and the empty set attains $b=0$. Since the query itself has only $r$ positions, the largest possible overlap satisfies
\[
    b\le b^*\le\max\{b,\min(m,r)\}.
\]
If the local test rejects $L_{m,b+1}$, Equation~\eqref{eq:least-favourable-family} rules out every size-$m$ set with overlap greater than $b$, so we can decrease $m$. Otherwise we update $b$ to the overlap of this set, $|L_{m,b+1}\cap R|\ge b+1$. Both updates preserve these bounds on $b^*$. At termination, either $b=r$ or $m\le b$; no remaining set can improve the attained overlap, and the displayed bounds give $b=b^*$. Returning $1-b/r$ therefore gives the closed-testing proportion bound.

Finally, each test decreases the nonnegative integer $m-b$ by at least one, from an initial value of $n$, so there are at most $n$ evaluations. Sorting the evidence scores and, for sign-based tests, their magnitudes costs $O(n\log n)$. Given these orders, constructing a representative and evaluating its local test each take $O(n)$ time for the tests listed above. The total cost is therefore $O(n\log n+n^2)=O(n^2)$, excluding fitting and calibration.
\end{proof}

\FloatBarrier

\section{Technical Details}\label{app:technical-details}

We describe how we fit the working models for Gumbel--max scores and calibrate the local tests. We then give the corresponding scores and working models for SynthID-Text and describe how we search for passages under a reading budget in the database-recovery review.

\subsection{Working-model specification and estimation}\label{app:working-models}

For Gumbel--max watermarking, the HMM and change-point models use the same emission family for the score magnitudes. We derive this family and describe how each model estimates the weights in Equation~\eqref{eq:sign-flip-score}.

\subsubsection{Emission distribution}\label{app:gamma-mixture}

For Gumbel--max watermarking, $U_j=F_0(Y_j)=Y_j$ is uniform under the null. Under the alternative, for a fixed next-token distribution $P_j$, differentiating Equation~\eqref{eq:wm-gumbel-alternative} gives the pivot density
\begin{equation}\label{eq:gumbel-mixture-density}
    h_{P_j}(u)=\sum_{w:P_{j,w}>0}u^{1/P_{j,w}-1}
    =\int_{[1,\infty)}u^{x-1}\,N_{P_j}(dx),
    \qquad N_{P_j}=\sum_{w:P_{j,w}>0}\delta_{1/P_{j,w}}.
\end{equation}
Here $\delta_x$ denotes a unit point mass at $x$. Each summand is a $\mathrm{Beta}(1/P_{j,w},1)$ density multiplied by its mixture weight $P_{j,w}$. Thus $h_{P_j}$ is a probability density, although $N_{P_j}$ is a counting measure. The verifier generally does not know the next-token probabilities, which may vary across positions. We replace the counting measure in Equation~\eqref{eq:gumbel-mixture-density} with a continuous mixing distribution and normalize the kernel. Taking a limit then gives a working family with one signal-strength parameter.

For a probability law $G$ on $[1,\infty)$, replacing the counting measure and normalizing gives
\[
    h_G(u)=\frac{\int u^{x-1}G(dx)}{\int x^{-1}G(dx)}
          =\frac{\E_G(u^{X-1})}{\E_G(X^{-1})}.
\]
For $X=1+S$ with $S\sim\mathrm{Gamma}(k,\mathrm{rate}=r)$, $\E(u^S)=\{r/(r-\log u)\}^k$. Canceling $r^k$ gives
\[
    h_{k,r}(u)
    =\frac{(r-\log u)^{-k}}
           {\int_0^1(r-\log v)^{-k}\,dv},
    \qquad 0<u<1.
\]
For $0<k<1$, the substitution $t=-\log u$ gives
\[
    \int_0^1(r-\log u)^{-k}\,du
    \longrightarrow \int_0^\infty t^{-k}e^{-t}\,dt
    =\Gamma(1-k)
    \qquad\text{as }r\downarrow0.
\]
Setting $\nu=1-k$ and including the uniform endpoint $\nu=1$ yields the working family in Equation~\eqref{eq:gamma-pivot-emission}. The verifier can fit this family without knowing the next-token probabilities. A poor fit can affect power. Theorem~\ref{thm:sact-validity} guarantees simultaneous coverage under the conditional null model even if the working family fits poorly.

The Gamma working family has pivot density $h_\nu(u)=(-\log u)^{\nu-1}/\Gamma(\nu)$ for $0<u<1$. To obtain the density of $A_j=|Z_j|$ at $a>0$, we add the densities of $Z_j$ at $a$ and $-a$. Under the working alternative the signed score has density $\phi(z)h_\nu\{\Phi(z)\}$. The two magnitude densities are
\begin{equation}\label{eq:hmm-working-emissions}
    g_0(a)=2\phi(a),\qquad
    g_1(a;\nu)=\phi(a)\bigl[h_\nu\{\Phi(a)\}
                         +h_\nu\{\Phi(-a)\}\bigr],
    \qquad a\geq0.
\end{equation}
Here $\phi$ denotes the standard normal density. To fit either structural model, we use the emission log likelihood ratio
\[
    \ell_j(\nu)=\log\frac{g_1(A_j;\nu)}{g_0(A_j)}
       =\log\left[\frac{h_\nu\{\Phi(A_j)\}+h_\nu\{\Phi(-A_j)\}}{2}\right].
\]
At ineligible positions, we set this log likelihood ratio to zero, so the missing observation does not favor either working state. These positions remain in the sequence and in query denominators but are excluded from the sign scans.

\subsubsection{Hidden Markov model}\label{app:hmm-fitting}

We start the Markov chain in its stationary distribution $(1-\pi,\pi)$. Here $\pi$ is the stationary probability of the active state, and $d$ is the mean length of an active run. Its transition matrix is
\begin{equation}\label{eq:hmm-transition}
    \mathbf P(\pi,d)
    =\begin{pmatrix}
       1-\dfrac{\pi}{d(1-\pi)} & \dfrac{\pi}{d(1-\pi)}\\[6pt]
       \dfrac{1}{d} & 1-\dfrac{1}{d}
     \end{pmatrix},
\end{equation}
with $d>1$ and $0<\pi<d/(d+1)$, so all transition probabilities lie strictly between zero and one. The magnitude emissions are given in Equation~\eqref{eq:hmm-working-emissions}.

We regularize the transition parameters with
\begin{equation}\label{eq:hmm-transition-prior}
    \operatorname{logit}(\pi)
       \sim N\{\operatorname{logit}(0.25),1\},
    \qquad
    \log d\sim N\{\log 32,(\log 4)^2\},
\end{equation}
independently before truncation to the valid region. We integrate over this continuous transition prior using a tensor-product Gauss--Hermite quadrature rule.

We estimate $\nu\in[0.001,1]$ by maximizing the marginal likelihood of the observed magnitudes. We sum over state paths using forward--backward recursions and integrate over the two transition parameters using seventh-order quadrature in each coordinate. We discard quadrature nodes outside the parameter space and renormalize the remaining prior weights. We run L-BFGS-B on $\log\nu$ from several starting values and keep the fit with the highest likelihood. To compute the gradient, we weight the emission derivatives by the posterior state probabilities. With $B$ likelihood evaluations and $Q$ retained tensor-product nodes, fitting costs $O(BQn)$.

Let $(\pi_b,d_b)$, $b=1,\ldots,Q$, be the quadrature nodes and let $\widehat\omega_b$ be their fitted posterior weights. Averaging over the nodes gives the active-state probability used in Equation~\eqref{eq:sign-flip-score},
\begin{equation}\label{eq:hmm-fitted-probability}
    q_j=\sum_{b=1}^Q\widehat\omega_b
        \Pr_{\widehat\nu,\pi_b,d_b}
        (G_j=1\mid A_{1:n}).
\end{equation}
The transition nodes lie in the interior and the emission densities are positive, so $q_j\in(0,1)$. Null score magnitudes $q_jA_j$ are therefore positive almost surely under the continuous pivot law.

\subsubsection{Change-point model}\label{app:cp-fitting}

Let $K(G)$ count the nonempty active runs in a binary path of length $n$. We use a Poisson$(\lambda)$ prior truncated to $0\leq K\leq K_{\max}$, with $K_{\max}=32\wedge\lfloor(n+1)/2\rfloor$, and assign equal probability to all paths with the same $K$. There are $\binom{n+1}{2K}$ such paths, including runs that touch a document boundary. Thus
\[
    \Pr(G=g)=\frac{p_{K(g)}}{\binom{n+1}{2K(g)}},
    \qquad
    p_k=\frac{\lambda^k/k!}{\sum_{h=0}^{K_{\max}}\lambda^h/h!}.
\]
The all-null path has $K=0$. We use $\lambda=2$ in the main CP analysis and repeat the analysis with $\lambda=0.5$ to assess sensitivity to the prior.

For emission log ratios $\ell_j(\nu)$, let $F_j(s,k)$ sum the likelihood ratios of paths ending in state $s$ with $k$ active runs. Initialize $F_0(0,0)=1$ and all other entries to zero. The recursion is
\[
\begin{aligned}
 F_j(0,k)&=F_{j-1}(0,k)+F_{j-1}(1,k),\\
 F_j(1,k)&=e^{\ell_j(\nu)}\{F_{j-1}(1,k)+F_{j-1}(0,k-1)\}.
\end{aligned}
\]
Terms with invalid indices are zero. The marginal likelihood ratio is
\[
    L_{\mathrm{CP}}(\nu)=\sum_{k=0}^{K_{\max}}
       \frac{p_k}{\binom{n+1}{2k}}\{F_n(0,k)+F_n(1,k)\}.
\]
Backward recursions then give the posterior probabilities of interval membership and the posterior distribution of $K$. We evaluate the recursions on the log scale and maximize over $\nu$ using the same parameter bounds and optimization method as the HMM. Each objective and gradient evaluation costs $O(nK_{\max})$. Both models supply $q_j$ to the score in Equation~\eqref{eq:sign-flip-score}.

\subsection{Local tests and calibration}\label{app:sact-calibration}

We group the procedures used in Section~\ref{sec:numerical} into sign-based tests, tests based on $p$-values, and CountScan's interval-based construction. All use the same verification statistics and candidate query sets. For a tested set $I$, write $m=|I|$.

\subsubsection{Sign-based tests}\label{app:sign-tests}
Using the fitted scores in Equation~\eqref{eq:sign-flip-score}, we define local tests calibrated by the conditional sign-flip law in Section~\ref{subsec:score-symmetry}. Within $I$, order the scores by decreasing magnitude. Let $P_{I,k}$ be the number of positive signs among the first $k$ scores.

\paragraph*{Power-divergence scans.}
At each rank $k$, we compare the fraction of positive signs, $P_{I,k}/k$, with $1/2$. We use the power-divergence function
\[
 d_s(u,v)=\frac{u^sv^{1-s}+(1-u)^s(1-v)^{1-s}-1}{s(s-1)},
\]
with its continuous limit at $s=1$. We take the maximum over ranks at which positive signs are in the majority, giving
\begin{equation}\label{eq:power-divergence-sign-scan}
 T_{I,s}=\max_{k:P_{I,k}>k/2}k\,d_s(P_{I,k}/k,1/2),\qquad s>0.
\end{equation}
An empty maximum is zero. The $s=1$ case gives the Berk--Jones scan over the ranked signs \citep{berk1979gof}. At $s=2$, the statistic equals $\{\max(T_I,0)\}^2/2$, yielding SACT-$2$. We also evaluate $s=0.5,1,1.5,4$, while $s=3$ coincides with $s=2$. SACT-$Z$ and SACT-$q$ retain the $s=2$ test but use $Z_j$ and $\operatorname{sign}(Z_j)q_j$, respectively.

\paragraph*{Barber--Cand\`es procedure.}

The scan based on the Barber--Cand\`es procedure \citep{barber2015knockoffs} is
\begin{equation}\label{eq:bc-sign-scan}
 T_{I,\mathrm{BC}}=\max_{1\le k\le m}\frac{P_{I,k}}{1+k-P_{I,k}}.
\end{equation}
For each intersection size $m$, we use the smallest attainable cutoff with null upper-tail probability at most $0.05$, or $+\infty$ if none exists. We compute this probability recursively, counting positive signs as we add each of the $m$ independent fair signs in turn. The resulting closed-testing bounds use the reference-family algorithms in Appendices~\ref{app:proofs-computation} and~\ref{app:scan-extensions}. Ties are resolved by position, independently of the signs. For computation, we assign distinct magnitudes consistent with this order.

\paragraph*{KJI.}
KJI tests the same cumulative counts $P_{I,k}$ of positive signs at prespecified checkpoints \citep{li2024knockofftdp}, giving
\[
 \phi_I^{\mathrm{KJI}}=\ind\{\exists v:\ P_{I,\min(m,k_v+v-1)}\ge k_v\},
\]
with thresholds fixed across intersections. Grid A uses $v_i=i$. Grid B starts at $v_1=1$ and uses $v_i=\lfloor i^2/2\rfloor$ for $i\ge2$. Grid C uses the Fibonacci sequence starting at $(1,2)$, and grid D uses $v_i=2^{i-1}$. In each grid we retain only checkpoints with $v_i\le\lfloor n/3\rfloor$. Using Algorithm~2 of \citet{li2024knockofftdp}, we simulate 200,000 fair-sign walks with a target crossing probability of $0.045$. We then compute the crossing probabilities exactly and adjust the thresholds to ensure level $0.05$. Unlike SACT, KJI uses prespecified checkpoints and thresholds shared across intersection sizes.

For computation, let $A_v$ contain the positive-score positions preceding the $v$th negative in decreasing magnitude order, or all positive positions if fewer than $v$ negatives occur. Precomputing these sets gives $d_\alpha^{\mathrm{KJI}}(R)=\max\{0,\max_v(|R\cap A_v|-k_v+1)\}$, an exact representation of full closure by Theorem~4.1 of \citet{li2024knockofftdp}.

\subsubsection{Tests based on p-values}
Using the power-divergence function $d_s$ defined above, GoF evaluates the sorted $p$-values in $I$ through
\[
 G_{I,s}=m\max_{k:k/m\ge p_{(k)}}d_s(k/m,p_{(k)}),\qquad s\in\{1,1.5,2\},
\]
including $k=m$. If $p_{(m)}=1$, this last rank contributes $d_s(1,1)=0$. GoF rejects when the observed statistic exceeds the $r$th largest of $M=200{,}000$ null calibration statistics, where $r=\lfloor0.05(M+1)\rfloor$. The calibration samples are independent of one another and of the experimental documents. The null rejection probability is at most $0.05$ after averaging over the random calibration samples. Simes rejects when $\min_k mp_{(k)}/k\le\alpha$. For both procedures, full closed-testing bounds are computed by Algorithm~\ref{alg:sact-querywise}, using evidence scores $-p_j$.

\subsubsection{CountScan}
We use the CountScan C1 evidence function with truncation 2.5 and hedging weight 0.5, a common log-evidence threshold $\log(n/\alpha)$, and bounds obtained by maximizing the sum of raw bounds over disjoint intervals contained in each query set \citep{chen2026countscan}. Here $n$ is the full document length before filtering. For a general set, we split its eligible positions into consecutive runs in retained document order, compute the optimal interval aggregation within each run by dynamic programming, and sum the results. The ranking determines which positions enter each set. Within the set, the scan uses document order. All methods use the same pivots, eligible positions, and rankings. The denominator for each proportion includes all tokens in the selected set. Comparisons concern retained-signal labels rather than CountScan's key-wide historical-match target. For SynthID-Text and Gaussian experiments, the same C1 construction uses conservative binomial and uniform-null $p$-values under the experiments' conditional null models.

\subsection{Extension to SynthID-Text watermarking}\label{app:synthid}

For SynthID-Text, we replace the Gumbel--max pivot and Gamma emission with a discrete verification score and binomial working distribution.

In the binary tournament of \citet{dathathri2024scalable}, the key assigns independent fair bits $g_{j,\ell}(w)$ to candidate tokens for each seeding context and layer. Repeated occurrences of the same candidate token share a bit. Each layer pairs candidates, favors the candidate with bit one, and breaks ties uniformly at random. Starting from $p^{(0)}_{j,w}=P_{j,w}$, the winner distribution satisfies
\begin{equation}\label{eq:wm-synthid-update}
\begin{aligned}
 r_{j,\ell}&=\sum_w p^{(\ell-1)}_{j,w}g_{j,\ell}(w),\\[3pt]
 p^{(\ell)}_{j,w}&=p^{(\ell-1)}_{j,w}\{1+g_{j,\ell}(w)-r_{j,\ell}\}.
\end{aligned}
\end{equation}
In our experiments we use $D=31$ layers and seed the watermark with the four preceding tokens. We sample each token from $p^{(D)}_j$.

Verification computes $S_j=\sum_{\ell=1}^D g_{j,\ell}(W_j)$. Under token--bit independence, $S_j\sim\mathrm{Binomial}(D,1/2)$. Define $Z_j=(2S_j-D)/\sqrt D$, $A_j=|Z_j|$, and $B_j=\max(S_j,D-S_j)$. Since $D$ is odd, the null scores are nonzero and their signs are fair conditional on their magnitudes. Under the conditional null model in Assumption~\ref{ass:null-pivots}, these signs are also independent, giving the sign-flip property required for calibration. We use the working alternative $\mathrm{Binomial}(D,\vartheta)$, $\vartheta\ge1/2$, whose masked masses are
\begin{equation}\label{eq:synthid-folded-emission}
 g_0(b)=2^{1-D}\binom Db,\qquad
 g_1(b;\vartheta)=\binom Db\{\vartheta^b(1-\vartheta)^{D-b}
                 +\vartheta^{D-b}(1-\vartheta)^b\}.
\end{equation}
Using these masked masses, we fit the HMM and CP models in Appendices~\ref{app:hmm-fitting} and~\ref{app:cp-fitting} by maximizing the magnitude-only marginal likelihood over $\operatorname{logit}\vartheta\in[0,8]$ to obtain $q_j$ and $X_j=q_jZ_j$. As for Gumbel--max, misspecification of the working model can affect power but does not affect validity under the conditional null model. Calibration uses the same fair-sign distribution, with ties resolved by position independently of the signs. The $p$-value methods use $p_j=\Pr\{\mathrm{Binomial}(D,1/2)\ge S_j\}$.

\subsection{Passage-selection heuristic}\label{app:passage-search}

The search in Section~\ref{subsec:database-exploratory} uses sentence boundaries, fitted scores, and exact bounds from Equation~\eqref{eq:database-interpolation}. For $m=1$, we enumerate all consecutive-sentence passages of at least 64 tokens within each budget and retain the one with the largest proportion lower bound. For $m>1$, we construct candidate passages, compare their unions, and refine the selected boundaries as follows.

\begin{enumerate}
    \item \emph{Construct a passage library.} Rank candidate passages $J$ by $\sum_{j\in J\cap\mathcal E}X_j/\sqrt{|J|}$. Group their lengths into bands with upper limits of 64, 96, 128, 192, 256, 384, 512, 768, 1,024, 1,536, and 2,048 tokens. Within each band, retain up to four passages in decreasing score order, skipping candidates whose intersection-to-union overlap with an already retained passage in that band exceeds $0.5$. To include weaker clusters, also form regions with $q_j\geq0.5$, merge gaps of at most 32 tokens, and retain regions containing at least 64 high-weight positions. For each such region and length band, add the highest-scoring passage with at least $80\%$ of its tokens in the region. Finally, include the optimal single-passage selections for each budget and remove duplicates.

    \item \emph{Evaluate passage combinations.} Enumerate pairs and triples from this library, retaining only separated passages with total length within budget. Evaluate each union jointly using Equation~\eqref{eq:database-interpolation}, and retain the set with the largest proportion lower bound for each $(B,m)$, including the single-passage selections as candidates.

    \item \emph{Refine the selected sets.} Make up to two passes over the selected passages. Working on one passage at a time, shift its endpoints by up to two sentence boundaries in either direction and retain the best feasible set. Also consider deleting a passage or adding one from the library when the passage limit allows. Every candidate must satisfy the length, separation, and budget constraints.
\end{enumerate}

For each budget and passage limit, we also consider sets evaluated under other limits if they meet the current constraints. This pool includes sets evaluated at $B=128$. We report results for $B\geq256$. To break ties, we prefer the larger certified count, then the shorter total length, then fewer passages, and finally earlier boundaries. The search uses only the observed text and watermark evidence. We compute the exact closed-testing bound for each set returned by the heuristic. Simultaneous coverage holds even if the search misses the optimal set.

\section{Supplementary Results}\label{app:sact-experiments}

We describe the simulation design and report empirical error rates before comparing detection power, token-level discovery, and passage localization under Gumbel--max and SynthID-Text watermarking. The final experiments examine how signal clustering, added background text, and the choice of working model and local test affect the bounds.

\subsection{Experimental design}\label{app:sact-data}

At $\alpha=0.05$, we compare SACT variants that use different local tests, scores, or working models. We also evaluate CountScan on the same documents, windows, and sets of top-ranked tokens. Appendix~\ref{app:sact-calibration} specifies the SACT local tests and comparison procedures.

\subsubsection{Document generation and signal labels}
We construct mixed documents using English web texts from the C4 realnewslike subset \citep{raffel2020t5} as background and passages generated by the LLM Qwen3.5-9B \citep{qwen2026qwen35}, with its thinking mode disabled. In each $n$-token document, we replace two intervals of $L$ consecutive token positions, separated by at least four tokens, with Gumbel--max watermarked passages. We generate these passages in document order, using the updated preceding text as context for the second passage. Seeding uses the four immediately preceding tokens. The prompt includes up to 128 tokens on either side of the passage. We begin generation with the 32 tokens immediately preceding the passage. The first generated token then has the same four-token seeding context as the position it replaces. We suppress special tokens and early termination until $L$ tokens have been generated. If a watermark-seeding context has already occurred, the next token is sampled from $P_j$ without watermarking. The generated passages are not edited.

We vary document length $n\in\{1024,2048\}$, passage length $L\in\{64,96,128,192,256\}$, and LLM sampling temperature $t\in\{0.8,1.0\}$. We use $t=1$ for the main comparisons and run 100 replicates per setting. For each document length, we exclude duplicate and previously used texts, then use the same background documents for all values of $L$ and $t$. A higher temperature flattens the next-token distribution and generally strengthens the Gumbel--max signal. Verification reconstructs $Y_j$, $p_j=1-Y_j$, and $Z_j=\Phi^{-1}(Y_j)$. A position receives $\xi_j=1$ only if it retains a valid watermarked token and matching generation context. Ineligible positions remain in query denominators but are excluded from tests.

The SynthID-Text experiments use the same C4 design, query windows, FDP targets, and permutation and dilution comparisons, with 100 all-null documents per length. We describe SynthID-Text generation and verification in Appendix~\ref{app:synthid}.

\subsubsection{Comparison procedures}
The default SACT procedure uses Gamma-HMM scores $q_jZ_j$ for Gumbel--max watermarking. SACT-$s$ varies the divergence index, and SACT-2-CP replaces the HMM with the CP working model. SACT-$Z$ and SACT-$q$ use $Z_j$ and $\varepsilon_jq_j$, respectively, with the $s=2$ test. SACT--BC and SACT--KJI change the local test while retaining the fitted scores. GoF-$s$ and Simes use pivotal $p$-values, while CountScan uses its C1 calibrator and interval-aggregation extension. Appendix~\ref{app:performance-comparisons} compares performance under both watermark mechanisms, and Appendix~\ref{app:sact-additional-results} examines the effects of structure adaptation.

\subsubsection{Structural experiments}
To examine the effect of added background, we start with 1,024-token documents containing one 512-token watermarked interval. We append the text following each document in its C4 source, increasing length to 1,280, 1,536, 1,792, 2,048, 2,560, 3,072, 3,584, and 4,096 while preserving the initial pivots and signal labels. For the permutation comparison, we permute the pivots, eligibility indicators, and signal labels together and then refit the model. We move windows of 128, 256, and 512 tokens in steps of 16 tokens. To assess score estimation when the generating model is known, we also simulate a Gaussian HMM with $n=1024$, stationary signal probability $0.25$, mean signal-run length 128, unit variance, null mean zero, and alternative means $0.75,1.25,2$. We compare fitted scores with oracle scores computed using the generating parameters.

\subsubsection{Computational evaluation}\label{app:computational-evaluation}
The timings in Table~\ref{tab:sact-central} cover 100 documents at $(n,L,t)=(2048,128,1)$. For each document, we time the preprocessing step and the evaluation of 1,000 randomly chosen sets of 256 tokens. All methods use the same sets. We use AMD EPYC 9654 processors with 64 single-threaded workers for the SACT, GoF, Simes, and KJI procedures. Sign and BC scans use the reference-family shortcut in Algorithm~\ref{alg:sact-bound}, GoF and Simes use Algorithm~\ref{alg:sact-querywise}, and KJI uses the reference-family representation described in Appendix~\ref{app:sign-tests}. For CountScan we use its exact interval-aggregation implementation with 128 single-threaded workers. We exclude the time spent loading data, generating query sets, fitting models, calibrating tests, and validating results. Intervals are pointwise 95\% Monte Carlo intervals across documents, except for the all-null table, which averages over keys.

\subsection{Empirical error control}\label{app:empirical-error-control}

\subsubsection{Gumbel--max and Gaussian experiments}
\begin{figure}[!tbp]
\centering
\includegraphics[width=.89\linewidth]{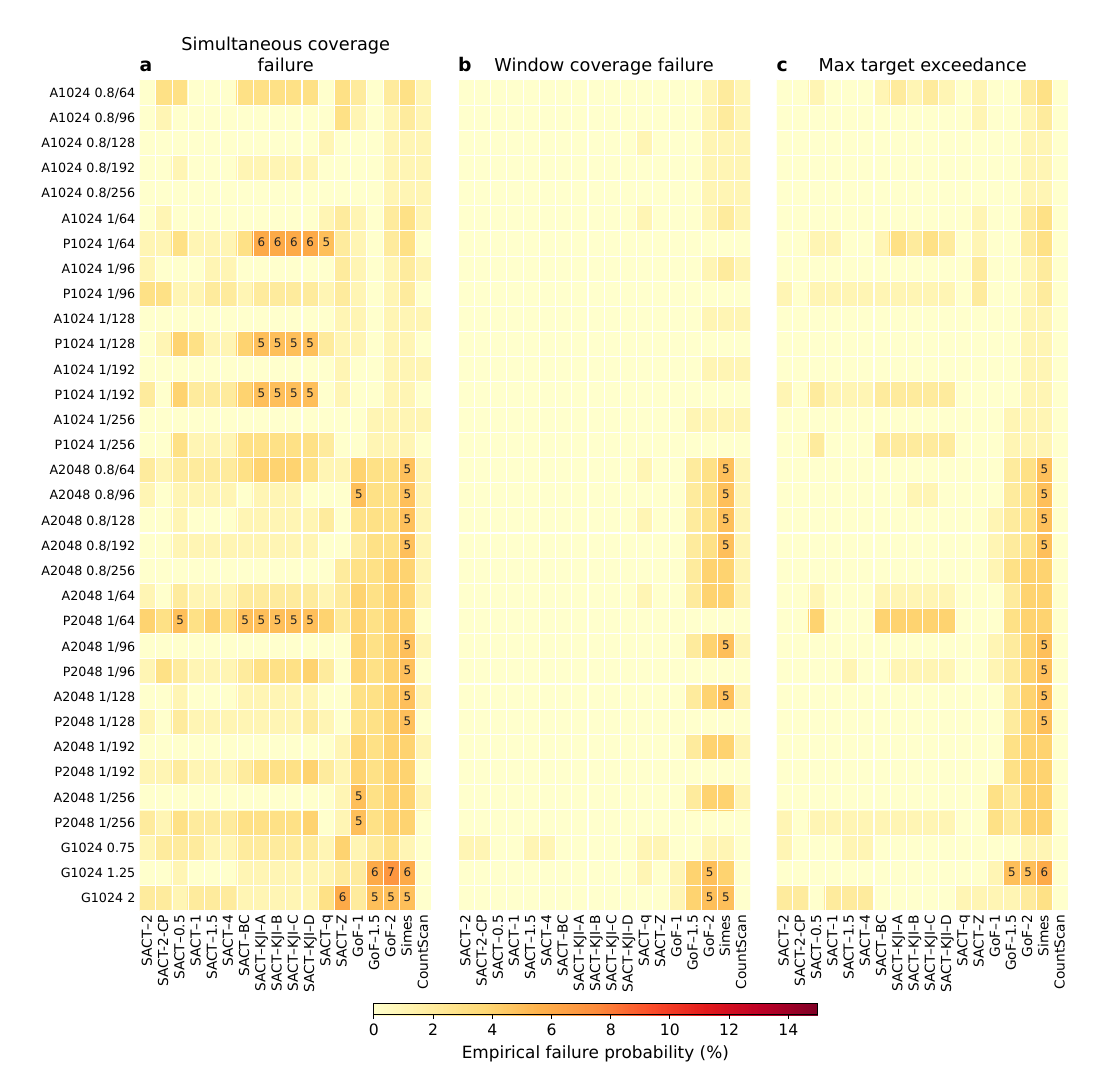}
\caption{Empirical error rates for all methods in 33 settings with signals (100 replicates per setting). Panel (a) shows how often simultaneous coverage fails, that is, $d_\alpha(\Nset)>0$ for the set $\Nset$ of true null positions. Panel (b) shows how often a bound exceeds the true count in at least one window. Panel (c) gives the largest of the six FDP exceedance rates calculated separately for two ranking paths and three targets. For CountScan, panel (a) checks the equivalent failure event over all intervals, and panel (c) uses its general-set bounds on the same paths. In the Gumbel--max rows, A denotes the original token order and P the permuted order. The labels give document length $n$, followed by temperature $t$ and passage length $L$ as $t/L$. Gaussian rows (G) give $n$ and the alternative mean. Numbers in the cells mark rates of at least 5\%.}
\label{fig:sact-app-coverage}
\end{figure}

Figure~\ref{fig:sact-app-coverage} reports three error rates. The first is the fraction of replicates in which simultaneous coverage fails. The second is the fraction in which the certified count exceeds the true signal count in at least one window. For the third, we estimate how often the realized FDP exceeds its target for each combination of ranking and target, then report the largest of these rates. All methods use the same windows and ranking paths for these comparisons. We retain all replicates, even when every lower bound is zero. For CountScan, we compare each interval bound with the retained-signal count in that interval. If every interval bound is correct, so is any sum over disjoint intervals. Across the 33 settings in Figure~\ref{fig:sact-app-coverage}, the largest observed rate for any of the three diagnostics was 7\% (7 of 100 replicates). For means, we use ordinary or paired Student-$t$ intervals. For probabilities, we use Clopper--Pearson intervals.

Table~\ref{tab:gamma-null} reports all-null rejection rates near the nominal 5\% level. We evaluate each cohort under 50 independent verification keys. For each key, we recompute the pivots and refit the models. Because all 50 keys are evaluated on the same texts, we compute a mean rejection rate for each key and use those 50 rates to form the confidence interval.

\begin{table}[!htbp]\centering
\renewcommand{\baselinestretch}{1}\small
\begin{tabular}{lrrll}\toprule
Null cohort & $n$ & Texts per key & HMM (\%) & CP, $\lambda=2$ (\%)\\\midrule
Primary & 1024 & 1,000 & 4.98 [4.77, 5.18] & 5.03 [4.81, 5.24]\\
Primary & 2048 & 1,000 & 4.98 [4.78, 5.18] & 5.00 [4.78, 5.22]\\
Separate cohort & 1024 & 2,000 & 4.97 [4.82, 5.11] & 5.02 [4.87, 5.17]\\
\bottomrule\end{tabular}
\caption{All-null rejection rates averaged over 50 independent verification keys for Gamma-HMM and Gamma-CP using SACT-$2$ at level 5\%. The same texts are evaluated under every key. Brackets are pointwise 95\% Student-$t$ intervals computed from the 50 key-level mean rates, conditional on these text cohorts.}
\label{tab:gamma-null}
\end{table}

\subsubsection{SynthID-Text experiments}
\begin{figure}[!tbp]
\centering
\includegraphics[width=\linewidth]{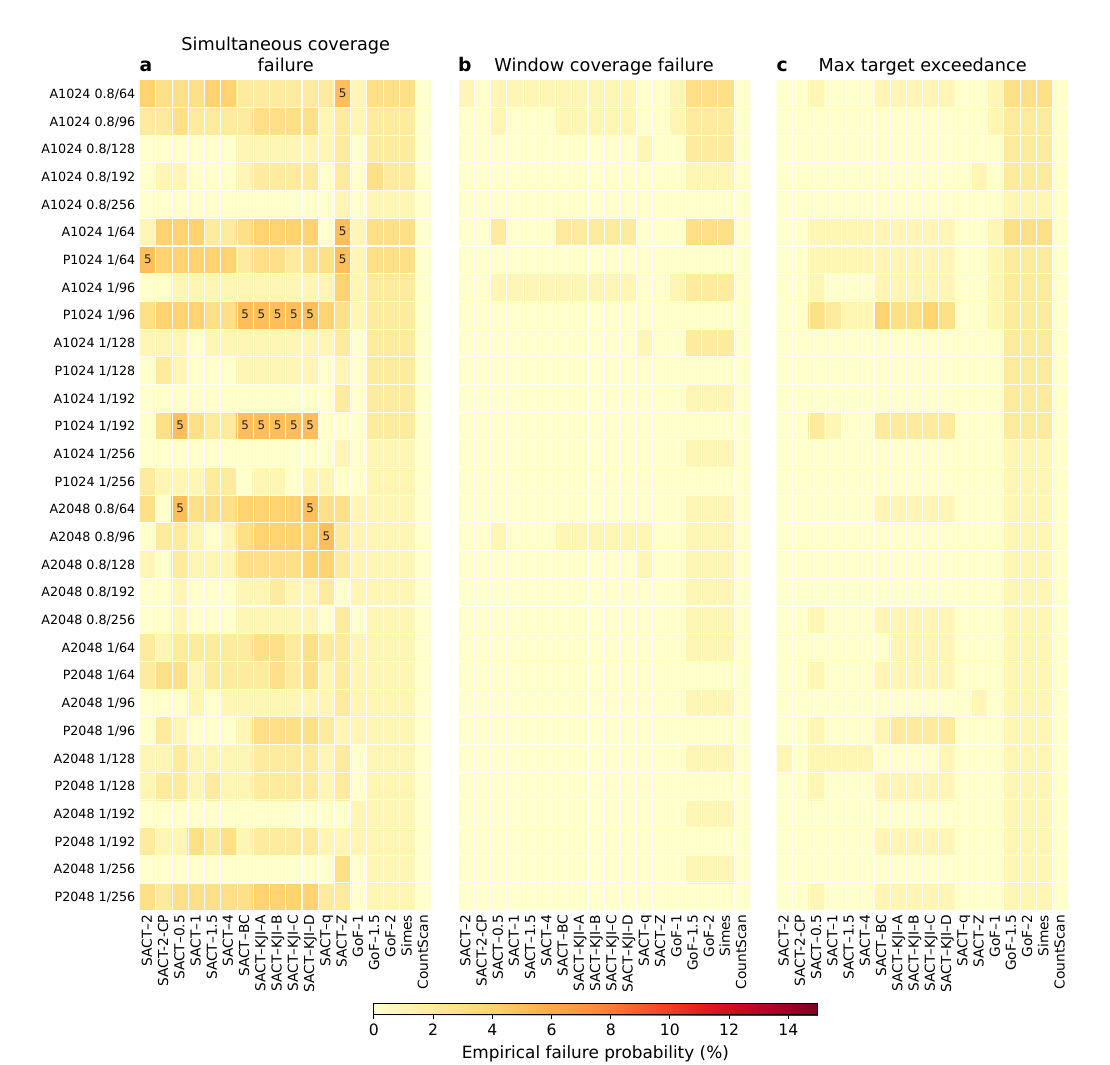}
\caption{SynthID-Text: error diagnostics across 30 intact or permuted signal settings and 18 methods, with 100 documents per setting. Panels report simultaneous coverage failure, overstatement in at least one window, and the largest of the six FDP exceedance rates for two rankings and three targets. For CountScan, the first panel checks the equivalent all-interval failure event and the third evaluates the same ranking paths using its general-set bounds. Row labels give the original (A) or permuted (P) token order, document length $n$, and the temperature and passage length as $t/L$. Rates of at least 5\% are annotated.}\label{fig:synthid-coverage}
\end{figure}

SACT-$2$ and SACT-2-CP have empirical simultaneous coverage of 99.25\% in the settings with unpermuted signals (Figure~\ref{fig:synthid-coverage}). Their all-null rejection rates are respectively 4\% and 3\% at $n=1024$, and 7\% and 6\% at $n=2048$. All four pointwise 95\% binomial intervals include the nominal level of 5\%.

\FloatBarrier

\subsection{Performance comparisons}\label{app:performance-comparisons}

We compare document-wide inference, token-level discovery, and passage localization under the two watermark mechanisms using the design in Appendix~\ref{app:sact-data}.

\subsubsection{Gumbel--max watermarking}\label{subsec:numerical-power}

\begin{figure}[!htbp]
\centering
\includegraphics[width=\linewidth]{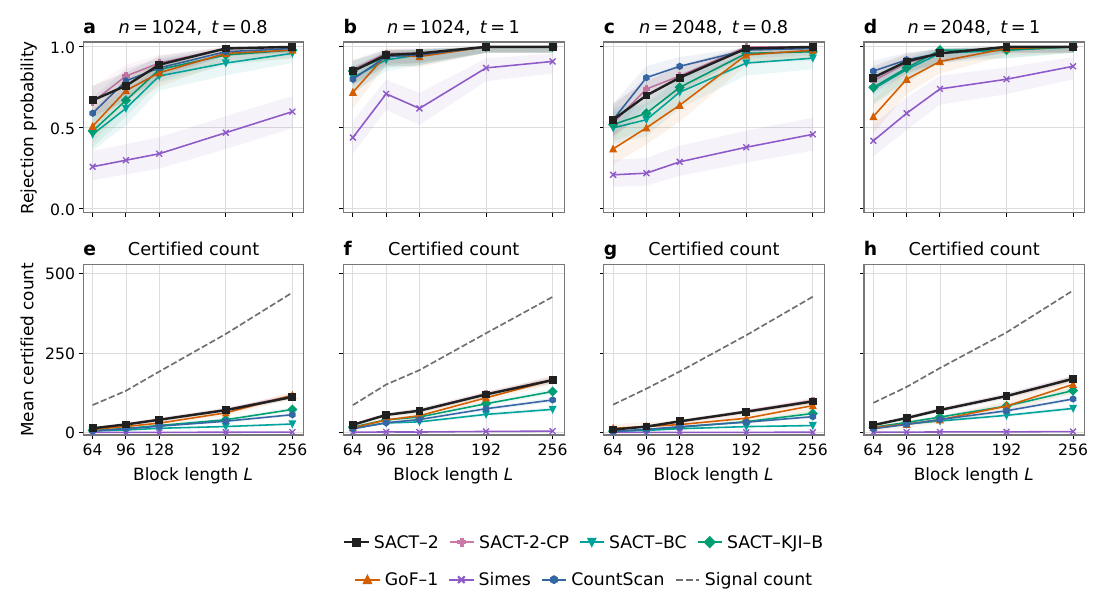}
\caption{Global rejection probability (top) and full-document certified counts (bottom) for C4 texts with two Gumbel--max watermarked passages. Columns show $(n,t)=(1024,0.8),(1024,1),(2048,0.8),(2048,1)$ across block lengths $L$. Dashed lines in the bottom row show the true signal counts. Each point is the mean over 100 replicates. Bands show 95\% Monte Carlo intervals.}
\label{fig:sact-global}
\end{figure}

\paragraph*{Global detection and proportion lower bounds.}
Figure~\ref{fig:sact-global} shows similar, generally high detection power for SACT variants and GoF-1, but larger certified counts for SACT-$2$ and SACT-2-CP than for GoF-1 and CountScan. We obtain the GoF-1 count bounds by applying closed testing to its local test.

\paragraph*{Token-level discovery.}
All methods query the same top-$k$ sets within each of three rankings: raw $Y_j$, HMM-based $X_j$, and CP-based $X_j$. As in Section~\ref{subsec:application-review}, we select the largest set meeting each FDP target $\gamma\in\{0.1,0.2,0.5\}$, separating the effects of the ranking and the local test.

\begin{figure}[tb]
\centering
\includegraphics[width=\linewidth]{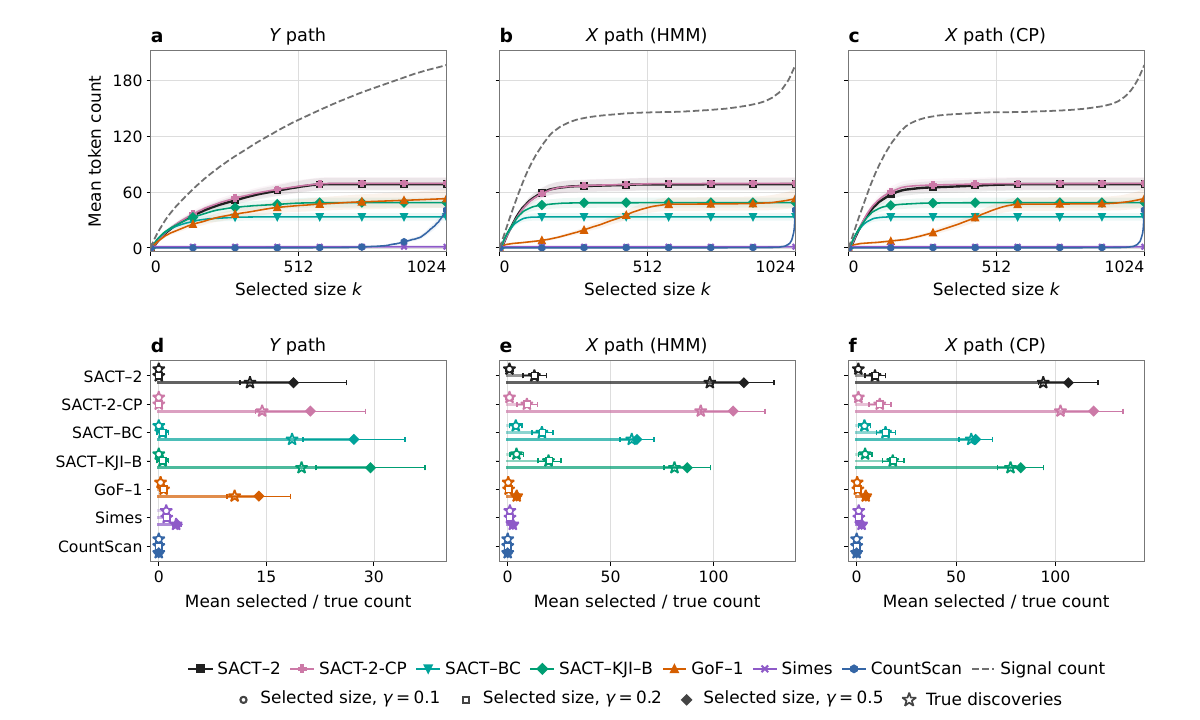}
\caption{Token-level discovery at $(n,L,t)=(1024,128,1)$ with two watermarked passages. Columns use the $Y$, HMM-based $X$, and CP-based $X$ rankings. The upper row shows certified counts along each path, with dashed curves giving signal counts. The lower row shows the number of selected tokens and the number of true discoveries at FDP targets $\gamma=0.1,0.2,0.5$ (circles, squares, and diamonds); hollow stars mark true discoveries. All methods query the same sets within each path. The plotted values are means over 100 replicates. Bands and error bars show 95\% intervals.}
\label{fig:sact-paths}
\end{figure}

Ranking by either working model places more signal tokens near the top of the list than ranking by $Y$ alone (Figure~\ref{fig:sact-paths}). BC and KJI--B give larger selections at stricter FDP targets, while at $\gamma=0.5$ SACT-$2$ and SACT-2-CP give the largest selections on their respective model-based paths. CountScan gives positive bounds only for larger top-$k$ sets. Even for those larger sets, the CountScan bounds are too small to meet any of the three FDP targets.

\begin{figure}[!tbp]
\centering
\includegraphics[width=\linewidth]{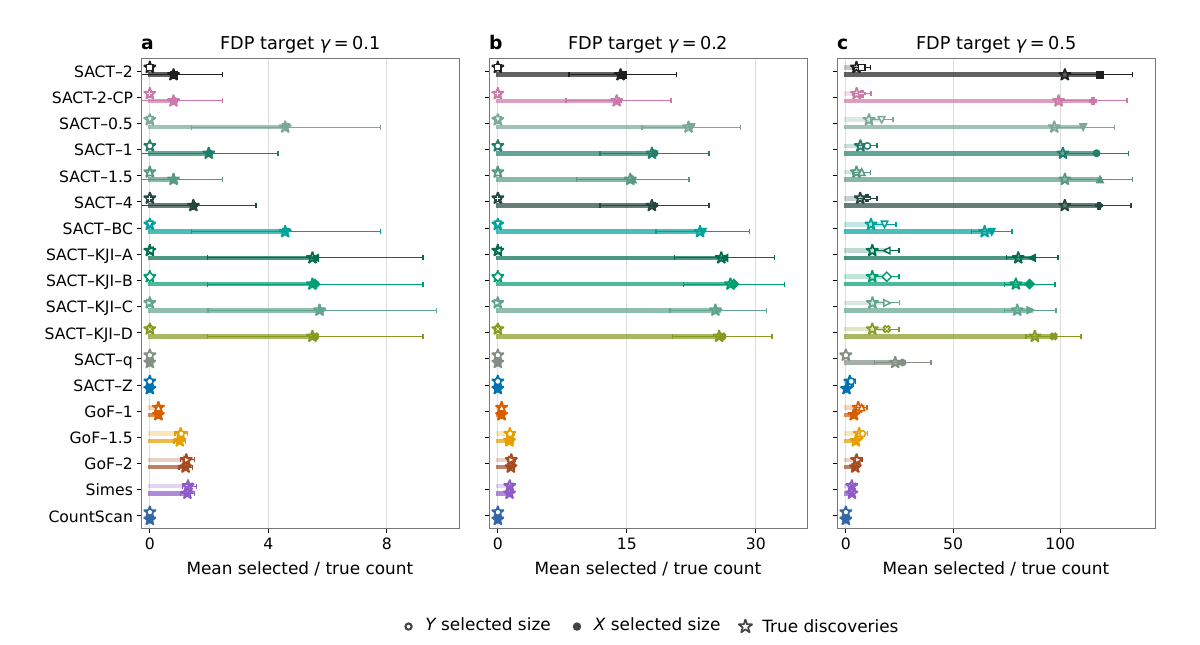}
\caption{Token selection under Gumbel--max watermarking at $(n,L,t)=(2048,128,1)$. Columns show FDP targets $0.1$, $0.2$, and $0.5$. Open and filled circles show the mean numbers of tokens selected using the $Y$ and HMM-based $X$ rankings, respectively. Hollow stars mark true discoveries. Bars are 95\% intervals across 100 replicates.}
\label{fig:sact-app-paths-2048}
\end{figure}

Figure~\ref{fig:sact-app-paths-2048} supplements the $n=1024$ path comparisons with selected sizes and true discoveries for all variants at $n=2048$.

\begin{figure}[tb]
\centering
\includegraphics[width=\linewidth]{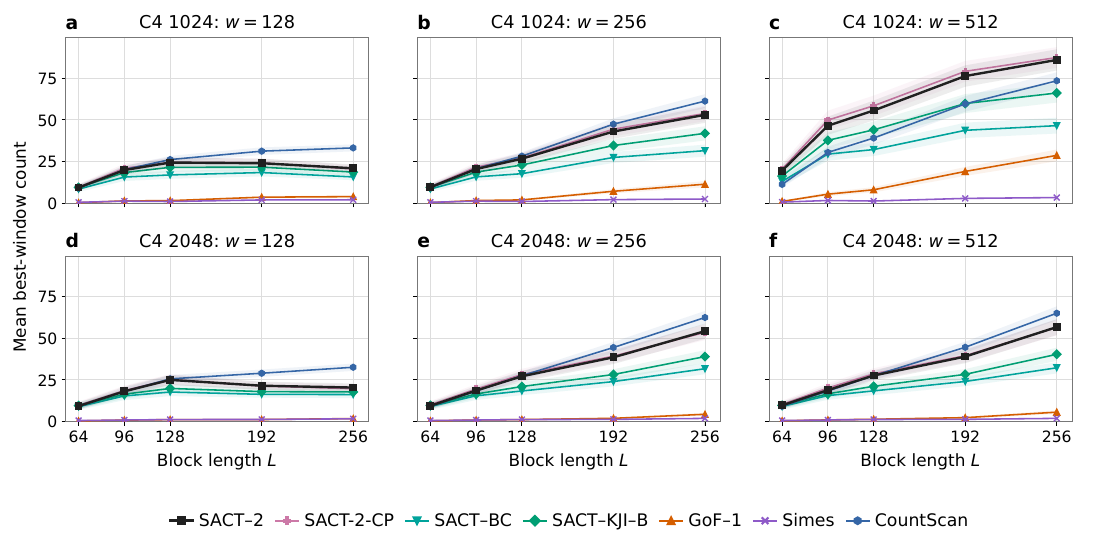}
\caption{Maximum certified counts over sliding windows at $t=1$. Rows correspond to document lengths and columns to window widths. The windows advance in steps of 16 tokens. Each method selects its window using the bound alone. Each point is the mean over 100 replicates. Bands show 95\% Monte Carlo intervals.}
\label{fig:sact-localization}
\end{figure}

\paragraph*{Passage-level localization.}
For each window width, each method selects the window with the largest certified count from the common grid. Structure-adaptive methods give larger counts than GoF-1 and Simes, while CountScan often gives larger certified counts for short windows (Figure~\ref{fig:sact-localization}).

\FloatBarrier

\subsubsection{SynthID-Text watermarking}\label{app:synthid-results}

Using the verification statistics and working model in Appendix~\ref{app:synthid}, we repeat the comparisons in Appendix~\ref{app:sact-data} under SynthID-Text watermarking. Appendix~\ref{app:empirical-error-control} reports the corresponding error diagnostics.

\begin{figure}[!tbp]
\centering
\includegraphics[width=\linewidth]{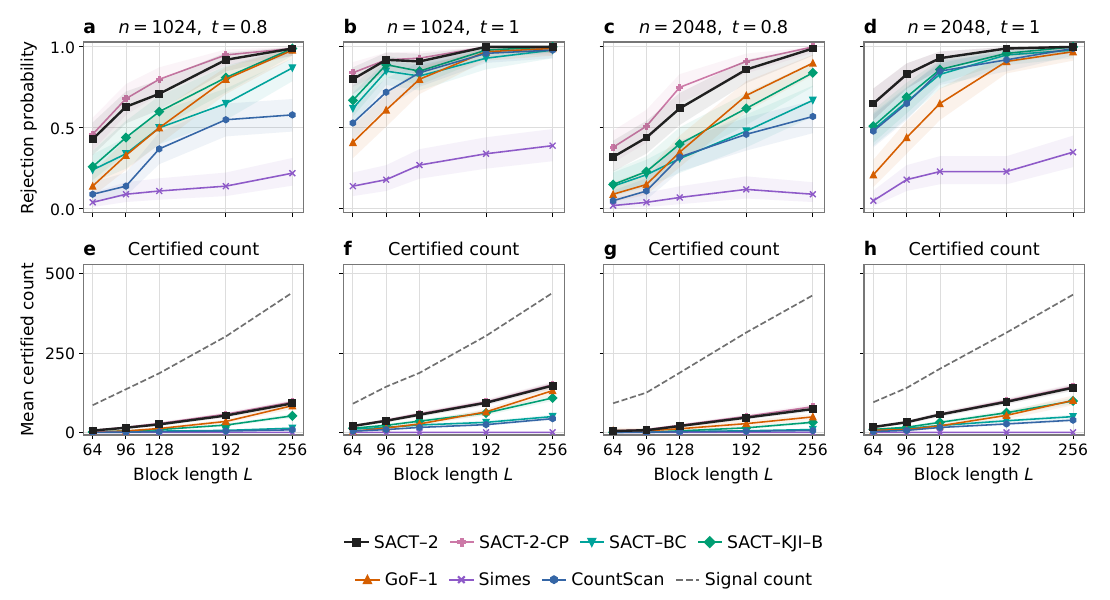}
\caption{Global detection power (top) and full-document certified counts (bottom) under SynthID-Text watermarking. Each column shows one combination of document length and sampling temperature. Each point is a mean over 100 documents. Dashed curves show signal counts and bands give pointwise 95\% Monte Carlo intervals. SACT-2-CP uses the CP prior with $\lambda=2$; the other structural methods use HMM weights.}\label{fig:synthid-global}
\end{figure}

\begin{figure}[!tbp]
\centering
\includegraphics[width=\linewidth]{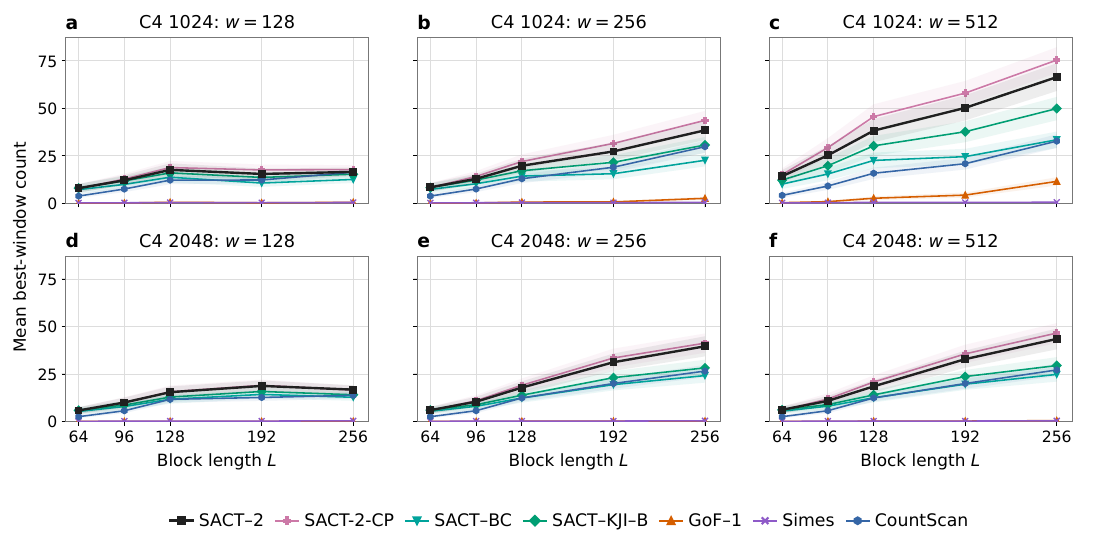}
\caption{Passage localization under SynthID-Text watermarking at $t=1$. For each width (128, 256, or 512 tokens), each method reports its largest certified count over windows spaced 16 tokens apart. Means and 95\% intervals use 100 documents per setting.}\label{fig:synthid-localization}
\end{figure}

\paragraph*{Global detection and passage localization.}
Across 20 settings with signals in their original order, SACT-$2$ and SACT-2-CP have mean detection powers of 79.70\% and 82.55\% and mean certified counts of 52.03 and 54.27 (Figure~\ref{fig:synthid-global}). CountScan detects a signal in 55.8\% of documents and certifies 10.84 signals on average. For each document, we also take the largest proportion bound over all 256-token windows. Averaged over the same 20 settings, this bound is 4.90\% for SACT-$2$, 5.54\% for SACT-2-CP, 3.23\% for CountScan, 0.29\% for SACT-$Z$, and 0.10\% for GoF-1. Figure~\ref{fig:synthid-localization} shows the corresponding window counts at $t=1$ for three window widths.

\begin{figure}[!tbp]
\centering
\includegraphics[width=\linewidth]{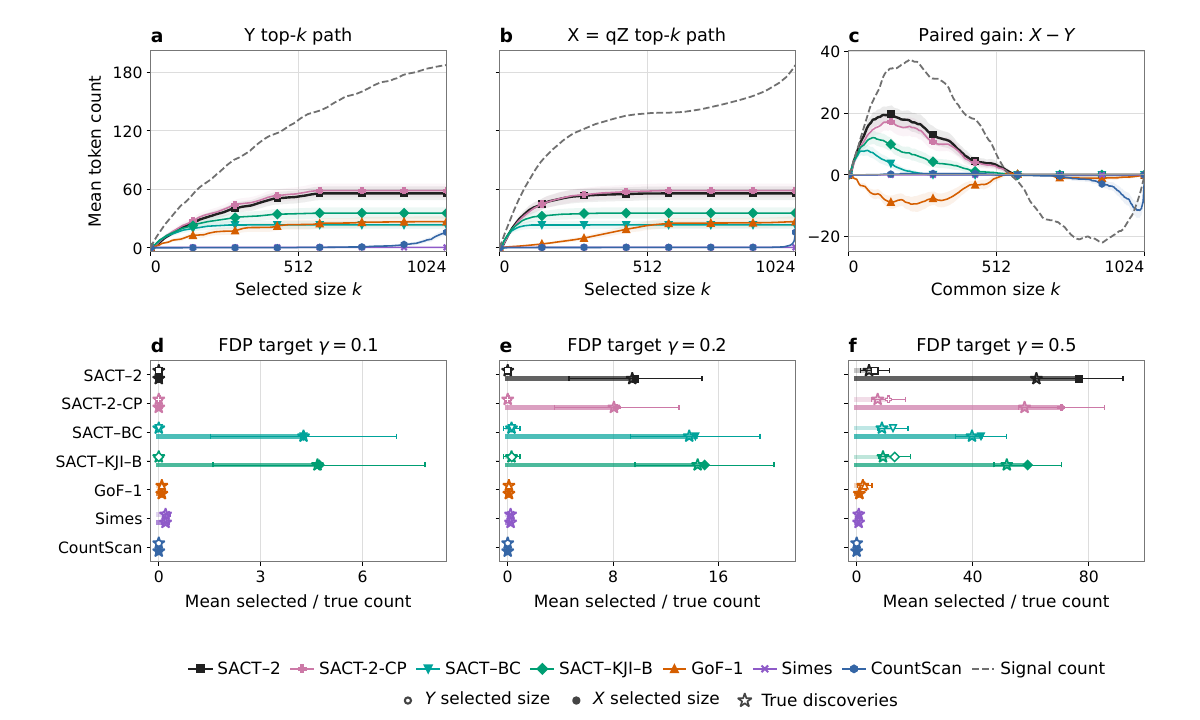}
\caption{SynthID-Text: token-level discovery at $(n,L,t)=(1024,128,1)$. Panels (a) and (b) show mean certified counts among the top $k$ tokens under the raw-count ($Y$) and HMM-based ($X$) rankings. Panel (c) shows the mean paired difference ($X$ minus $Y$) at each $k$. The dashed curves give the corresponding true signal counts in (a) and (b) and their difference in (c). In the bottom row, open and filled circles show the mean numbers selected under the $Y$ and $X$ rankings at each FDP target. Hollow stars mark the mean numbers of true discoveries. All methods use the same two rankings. Intervals are based on 100 documents.}\label{fig:synthid-selection}
\end{figure}

\paragraph*{Token-level discovery.}
All methods, including CountScan, evaluate the same top-$k$ sets under the raw-count and HMM-based rankings. At $(n,L,t)=(1024,128,1)$, CountScan selects no tokens at any of the three FDP targets (Figure~\ref{fig:synthid-selection}).

\FloatBarrier

\subsection{Effects of structure adaptation}\label{app:sact-additional-results}

We first examine signal clustering and dilution under both watermark mechanisms, then compare working models and local tests, including a Gaussian experiment with oracle scores.

\subsubsection{Signal clustering and background dilution}

\begin{figure}[tb]
\centering
\includegraphics[width=0.9\linewidth]{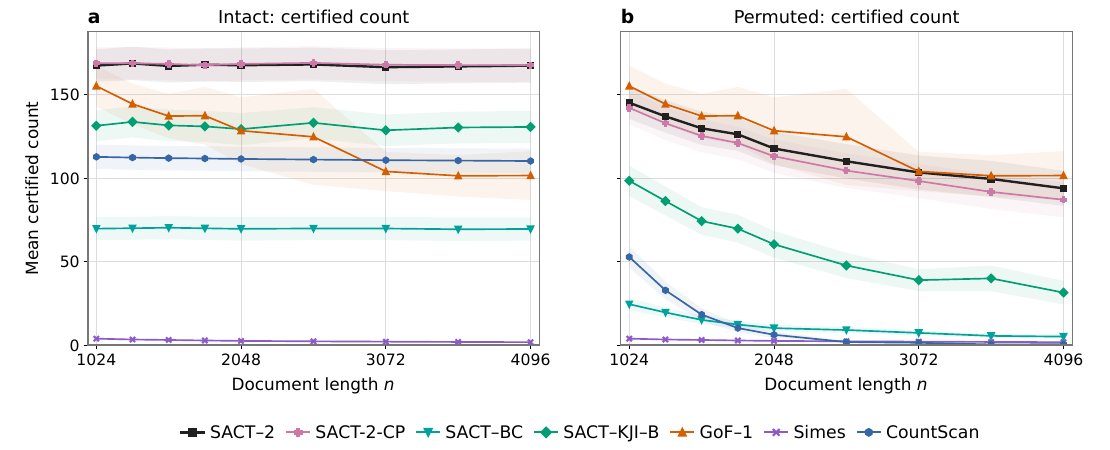}
\caption{Full-document certified counts under fixed-signal dilution at $t=1$, before and after joint permutation of the pivots, eligibility indicators, and signal labels. Each document initially has 1,024 tokens, including one 512-token watermarked interval. We lengthen each document by appending background text and keep the watermarked interval fixed. Bands are 95\% intervals across 100 paired documents.}
\label{fig:sact-dilution}
\end{figure}

\paragraph*{Gumbel--max watermarking.}
With a fixed 512-token watermarked interval, certified counts from structure-adaptive SACT variants and CountScan remain nearly stable as background is added, whereas GoF-1 counts decline (Figure~\ref{fig:sact-dilution}(a)). After joint permutation, the SACT counts decrease as the document grows, and SACT-$2$ gives counts similar to GoF-1 (Figure~\ref{fig:sact-dilution}(b)).

\begin{figure}[!tbp]
\centering
\includegraphics[width=.95\linewidth]{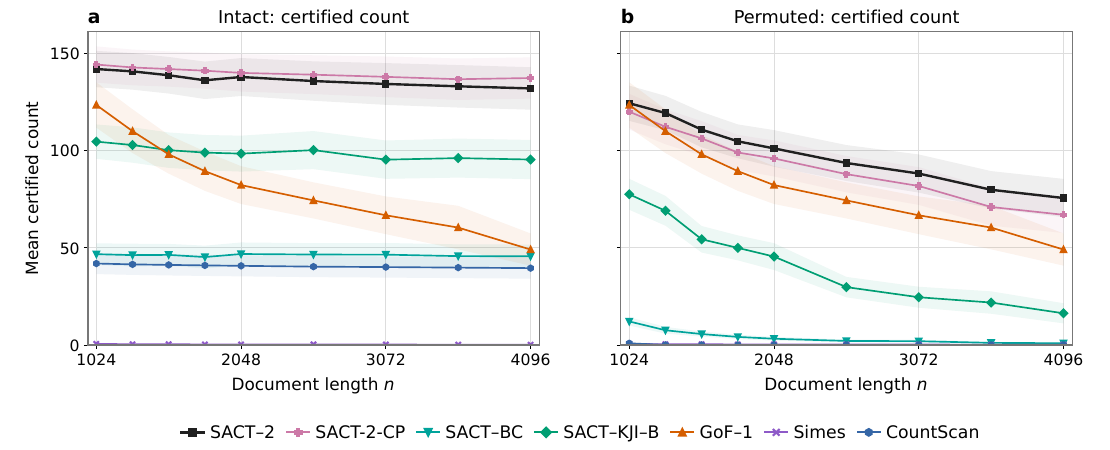}
\caption{Full-document certified counts as background text is added to a document containing one 512-token SynthID-Text watermarked passage at $t=1$. Panel (a) keeps the original token order. Panel (b) permutes the pivots, eligibility indicators, and signal labels together. Bands show 95\% intervals across 100 paired documents.}\label{fig:synthid-dilution}
\end{figure}

\paragraph*{SynthID-Text watermarking.}
For both SACT-$2$ variants, adding background text has less effect on the certified count when the watermarked passage stays intact than when the tokens are permuted (Figure~\ref{fig:synthid-dilution}).

\FloatBarrier

\subsubsection{Working models and local tests}

We compare HMM and CP weights using the same Gamma emission family and SACT-$2$ test. We first average the paired differences within each background document, then average over the 200 documents. For CP with $\lambda=2$, the mean gains over the HMM are 0.82 certified signals (95\% interval $[0.42,1.23]$), 0.28 percentage points in the maximum proportion lower bound over 256-token windows ($[0.19,0.38]$), and 0.45 percentage points in power ($[-0.38,1.28]$). The comparison of $\lambda=2$ and $\lambda=0.5$ gives similar results for detection power (90.25\% versus 89.45\%), mean certified count (67.75 versus 67.86), and mean window bound (8.69\% versus 8.28\%).

\begin{figure}[!tbp]
\centering
\includegraphics[width=\linewidth]{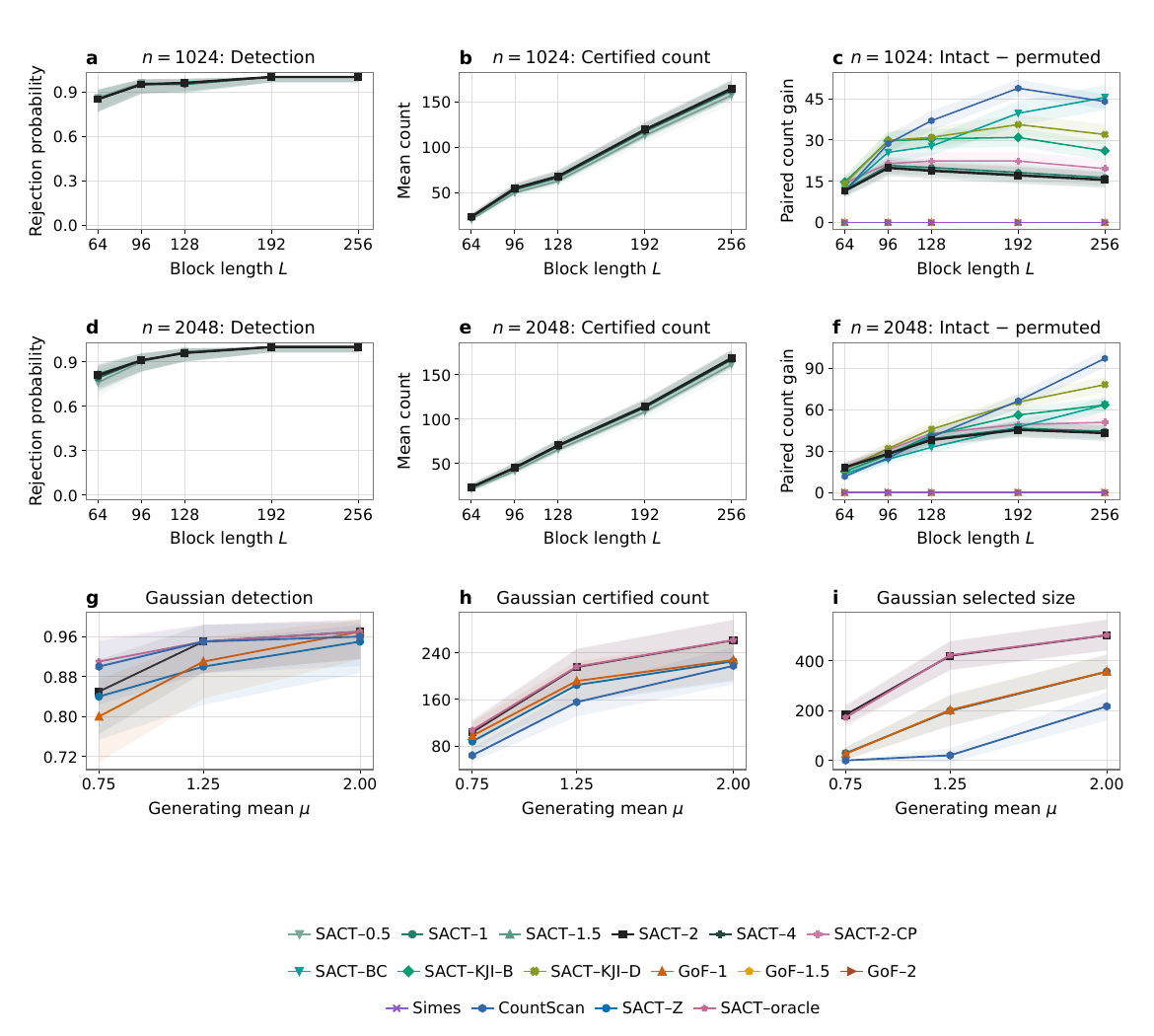}
\caption{Local-test and structural comparisons. The top two rows show Gumbel--max results at $t=1$ for $n=1024$ and $2048$, respectively. Columns show detection power, full-document certified counts, and the mean paired difference in certified counts (original order minus permuted order). The bottom row shows detection power, certified counts, and the number of selected positions at FDP target $0.5$ in the Gaussian experiment. CountScan is included for the paired count contrasts and all three Gaussian outcomes. The oracle uses the generating parameters to compute the scores for inference. The sets being evaluated still follow the ranking from the fitted $X$ scores. Bands show 95\% binomial intervals or 95\% $t$ intervals from ordinary or paired comparisons.}
\label{fig:sact-app-structure}
\end{figure}

\FloatBarrier

\end{document}